\documentclass[aps,prd,preprintnumbers,nofootinbib,superscriptaddress,twocolumn]{revtex4} 
\usepackage{ae}
\usepackage{bm} 
\usepackage{xcolor}
\usepackage{amssymb}
\usepackage{amsmath}
\usepackage{amsfonts}
\usepackage{graphicx}
\usepackage{slashed}
\usepackage{wasysym}
\usepackage{setspace}
\usepackage{ulem}
\usepackage{multirow}
\usepackage{footnote}
\usepackage{float}
\usepackage{diagbox} 
\usepackage{ulem} 
\usepackage{adjustbox} 
\usepackage{hyperref} 

\newcommand{\stkout}[1]{\ifmmode\text{\sout{\ensuremath{#1}}}\else\sout{#1}\fi}

\newcommand{\Eqref}[1]{Eq.~\eqref{#1}}
\newcommand{\Figref}[1]{Fig.~\ref{#1}}

\definecolor{shiraz}{rgb}{0.6,0,0.4}

\allowdisplaybreaks

\begin{document}

\setlength{\unitlength}{1mm}
\title{All-optical Quantum Vacuum Signals in Two-Beam Collision}

\author{Holger Gies}\email{holger.gies@uni-jena.de}
\affiliation{Helmholtz-Institut Jena, Fr\"obelstieg 3, 07743 Jena, Germany}
\affiliation{GSI Helmholtzzentrum f\"ur Schwerionenforschung, Planckstra\ss e 1, 64291 Darmstadt, Germany}
\affiliation{Theoretisch-Physikalisches Institut, Abbe Center of Photonics, \\ Friedrich-Schiller-Universit\"at Jena, Max-Wien-Platz 1, 07743 Jena, Germany}
\author{Felix Karbstein}\email{felix.karbstein@uni-jena.de}
\affiliation{Helmholtz-Institut Jena, Fr\"obelstieg 3, 07743 Jena, Germany}
\affiliation{GSI Helmholtzzentrum f\"ur Schwerionenforschung, Planckstra\ss e 1, 64291 Darmstadt, Germany}
\affiliation{Theoretisch-Physikalisches Institut, Abbe Center of Photonics, \\ Friedrich-Schiller-Universit\"at Jena, Max-Wien-Platz 1, 07743 Jena, Germany}
\author{Leonhard Klar}\email{leonhard.klar@uni-jena.de}
\affiliation{Helmholtz-Institut Jena, Fr\"obelstieg 3, 07743 Jena, Germany}
\affiliation{GSI Helmholtzzentrum f\"ur Schwerionenforschung, Planckstra\ss e 1, 64291 Darmstadt, Germany}
\affiliation{Theoretisch-Physikalisches Institut, Abbe Center of Photonics, \\ Friedrich-Schiller-Universit\"at Jena, Max-Wien-Platz 1, 07743 Jena, Germany}

\date{\today}

\begin{abstract}
The fundamental theory of quantum electrodynamics predicts the vacuum to resemble a polarizable medium.
This gives rise to effective nonlinear interactions between electromagnetic fields and light-by-light scattering phenomena.
We study the collision of two optical laser pulses in a pump-probe setup using beams with circular and elliptic cross section and estimate the number of discernible signal photons induced by quantum vacuum nonlinearities.
In this analysis we study strategies to optimize the quantum vacuum signal discernible from the background of the driving lasers.
One of the main results is that the collision of two maximally focused lasers does not lead to the best discernible signal.
Instead, widening the focus typically improves the signal to background separation in the far field.

\end{abstract}

\maketitle

\section{Introduction}\label{sec:intro}

As the quantitatively best verified quantum field theory within the Standard 
Model of particle physics, quantum electrodynamics (QED) still offers parameter 
regimes untested in experiment.
The present study is devoted to the regime pioneered by Heisenberg and Euler introducing an effective Lagrangian \cite{Heisenberg:1935qt}, which encodes the effects of QED vacuum fluctuations in effective nonlinear interactions between macroscopic electromagnetic fields.
An in-depth study of this regime does not only expand our understanding 
of this fundamental theory but could also shed light on physics beyond the 
Standard Model (BSM), e.g., \cite{Born34,Gies:2006ca,Ahlers:2007qf,Gies:2008wv,Jaeckel:2010ni,Baker:2013zta,Davila:2013wba,Villalba-Chavez:2016hxw,Rebhan:2017zdx,Agrawal:2021dbo}.
 
The quantum vacuum can be interpreted as a quantum state characterized by fluctuations of particle and antiparticle pairs on short space and time scales.
In the case of QED, these are electrons and positrons interacting with photons.
As fields couple to charges these vacuum fluctuations induce the effective nonlinear interactions between external electrodynamic fields in the Heisenberg-Euler Lagrangian.
The nonlinear terms in the Lagrangian give rise to effects like photon-photon scattering or birefringence, which become sizable only at sufficiently high field strengths \cite{Euler:1935zz,Heinzl:2006xc}.
Thus, the quantum vacuum closely resembles a nonlinear medium in solid state physics.

In previous theoretical studies, many different signatures of the QED vacuum have already been analyzed; see the reviews~\cite{Dittrich:2000zu,Dunne:2004nc,Marklund:2006my,DiPiazza:2011tq,Battesti:2012hf,King:2015tba,Hill:2017,Inada:2017lop,Karbstein:2019oej,Schoeffel:2020svx,Fedotov:2022ely} and references therein. 
The main problem, however, is to distinguish between the small amount of signal photons and the comparably huge number of background photons of the driving electromagnetic fields. 

In this article, we study the prospects of inducing a discernible quantum vacuum signal in a two-beam pump-probe setup with optical ultrashort $\rm PW$ laser pulses, such as previously studied, e.g., by \cite{Tommasini:2010fb,King:2012aw,Gies:2017ygp, Robertson:2020nnc}.
A key parameter to enhance the signal is the choice of optimal beam waists.
Typically, tightly focused beams down to the diffraction limit are considered in order to maximize the involved field strength. While this does, in fact, generically maximize the scattering amplitudes, it does not necessarily optimize the signal-to-background ratio which is the relevant quantity to identify the quantum vacuum signature; cf. also \cite{Mosman:2021vua,Roso:2021hfo}. A crucial idea in this context is that the specific use of larger beam waists can scatter the quantum signal into lower-noise regions. This is, because the signal amplitude decreases only with a powerlaw for increasing waists, whereas the noise decreases exponentially in the relevant spacetime regions. 

In addition to studying beams with circular cross section, we also allow the probe pulse to have an elliptic focus cross section such as pioneered in the context of vacuum birefringence \cite{Karbstein:2015xra,Karbstein:2016lby}. 
We discuss the advantages of the circularly or elliptically focused probe beam settings for variable probe beam waists and determine the respective number of discernible signal photons mediated by quantum vacuum nonlinearites.

Our article is organized as follow. 
In section \ref{sec:form}, we briefly introduce the theoretical framework to reliably derive the number of signal photons arising from QED nonlinearites.  
Section \ref{sec:exp} specifies the experimental scenario, based on parameters available at state-of-the-art strong laser facilities.
In section \ref{sec:anal}, we identify the phase space regions where the number of signal photons dominates over the amount of background photons.
We then analyze the effects of different focusing on the yield for different collision angles between the pump and probe beams, both for circularly focused and elliptically focused probe beams.
Finally, we conclude with a summary of our results and a brief outlook in Sec.~\ref{sec:conc}. 

\section{Theoretical background: key to the quantum vacuum}\label{sec:form}

A convenient way to study the  signal photons induced by quantum vacuum nonlinearities is the \textit{vacuum emission picture} \cite{Galtsov:1971xm,Karbstein:2014fva}.
Here we assume the vacuum nonlinearites to be governed by the one-loop Heisenberg-Euler effective Lagrangian ${\cal L}_{\rm HE}$ \cite{Euler:1935zz,Heisenberg:1935qt,Weisskopf:1936kd,Karplus:1950zz,Schwinger:1951nm,Ritus:1975cf,Dittrich:1985yb,Dittrich:2000zu,Dunne:2004nc} and focus on the leading nonlinear correction to classical Maxwell theory.
In this limit, the relevant process can be illustrated by a closed fermion loop with four photon lines: three photon lines represent the driving laser pulses and the remaining line characterizes the photonic output of the nonlinear interaction.

In this section, we briefly summarize this approach and derive an expression for photonic signatures of quantum vacuum nonlinearities. We use the metric convention $g_{\mu\nu}=\mathrm{diag}(-,+,+,+)$ and Heaviside-Lorentz units with $\hbar=1=c$.

In order to induce a sizable signal, the driving electromagnetic fields should reach extremely high peak field strength values $E$ and $B$ at the feasible end of current technology.
On the other hand these fields are still weak compared to the field-strength scale constructed from the fundamental parameters of QED, i.e. $\{E,B\}\ll\frac{m_e^2}{e}$ with the elementary charge $e$ and the electron mass $m_e$.
In addition, we limit ourselves to optical and near-infrared frequencies with $\omega \ll m_e$.

It is convenient to introduce the field-strength invariants of electrodynamics, 
\begin{align}
\mathcal{F}&=\frac{1}{4} F^{\mu\nu} F_{\mu\nu}= \frac{1}{2}\left(\mathbf{B}^2-{\bf E}^2\right)\,, \nonumber \\
\mathcal{G}&=\frac{1}{4} \tilde{F}^{\mu\nu}F_{\mu\nu}= -\mathbf{B}\cdot{\bf E}\,, \label{eq:invariantFG}
\end{align}
where $\tilde{F}^{\mu\nu}=\frac{1}{2} \epsilon^{\mu\nu\alpha\beta}F_{\alpha\beta}$ is the dual field strength tensor. Then, the Heisenberg-Euler action ${\cal L}_{\rm HE} = {\cal L}_{\rm M} +  {\cal L}_{\rm int} $ consists of the classical Maxwell term $\mathcal{L}_{\text{M}}=-\mathcal{F}$ and the nonlinear interaction term induced by quantum fluctuations,
\begin{align}
 {\cal L}_{\rm int} &= \frac{m_e^4}{360\pi^2} \left(\frac{e}{m_e^2}\right)^4 \left(4 \mathcal{F}^2 + 7 \mathcal{G}^2 \right) + \dots \nonumber \\
 &=   \! \adjustimage{valign=c}{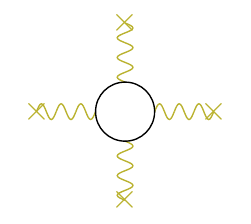}\! + \!\adjustimage{valign=c}{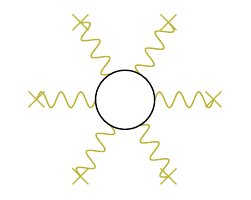}\! + \! \quad \dots 	\label{eq:Lint} 
\end{align}
Eqref{eq:Lint} is additionally depicted in terms of Feynman diagrams.
In the following, we limit ourselves to the lowest order photon-photon interaction, i.e. the contribution proportional to the square of the scalar invariants. 
Generalizations to higher orders are straightforward.

The number of quantum vacuum signal photons induced by the interactions of the driving fields can straightforwardly be derived within the vacuum emission picture. For this, we introduce the signal photon field $\hat{a}^{\mu}$ and treat it as an operator on the Fock space.
Replacing the electromagnetic fields according to $F^{\mu\nu}\to F^{\mu\nu}+\hat{f}^{\mu\nu}$ with the classical background field of the driving lasers $F^{\mu\nu}$ and the signal photon field $\hat{f}^{\mu\nu}=\partial^\mu\hat{a}^\nu-\partial^\nu\hat{a}^\mu$, we obtain the effective action $\Gamma_{\rm int}\left[\hat{a}\right]$ for the signal photons. 
This action mediates transitions between the vacuum state $\left|0\right>$ and the signal photon state $\left|\gamma_{(p)}(\mathbf{k})\right>$.
The induced signal photon of unit wave vector $\hat{\bf k}$ is characterized by a polarization vector $\epsilon^{\mu}_{\left(p\right)}\bigl(\hat{\bf k}\bigr)=(0,{\bf e}_{(p)})$ spanned by two transverse vectors ${\bf e}_{(p)}$ with $p\in\{1,2\}$, fulfilling $\hat{\bf k}\times{\bf e}_{(p)}={\bf e}_{(p+1)}$ and ${\bf e}_{(3)}=-{\bf e}_{(1)}$.

To evaluate the effective action $\Gamma_{\rm int}[\hat{a}(x)]=\int{\rm d}^4x\,{\cal L}_{\rm int}|_{F\to F+\hat{f}}$ we use the locally constant field approximation. This is quantitatively controlled by the spatiotemporal variation scales of the optical driving laser pulses compared to the time and space scale of QED; the Compton wavelength of an electron is $\lambdabar_C =3.86\times 10^{-13}\,{\rm m}$ and the Compton time is $\tau_{\rm C} = 1.29\times 10^{-21}\,{\rm s}$. 
Finally, the information about the signal photon distribution far outside the interaction region is encoded in the signal photon amplitude
\begin{align}
S_{(p)}\left(\mathbf{k}\right) &= \bigl\langle\gamma_{\left(p\right)}\left(\mathbf{k}\right)\bigr|\Gamma_{\text{int}}[\hat{a}(x)]\bigl|0\bigr\rangle\nonumber \\
& = \mathrm{i}\frac{\epsilon^{*\mu}_{\left(p\right)}\bigl(\hat{\bf k}\bigr)}{\sqrt{2k^0}} \int \mathrm{d}^4x\; \mathrm{e}^{\mathrm{i}k_{\alpha}x^{\alpha}} \nonumber \\
& \phantom{=} \times  \Bigl( k^{\nu} F_{\nu\mu} \frac{\partial \mathcal{L}_{\text{int}}}{\partial \mathcal{F}}  + k^{\nu} \tilde{F}_{\nu\mu} \frac{\partial \mathcal{L}_{\text{int}}}{\partial \mathcal{G}} \Bigr)\biggl|_{k^0={\rm k}} \nonumber \\
 &=  \adjustimage{valign=c}{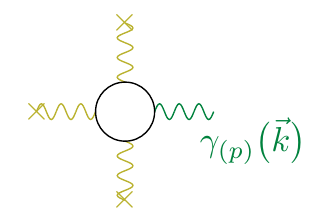} \,. \label{eq:S_p_Gamma}
\end{align}
Here, we use $*$ to denote the complex conjugation. Upon insertion of the corresponding derivatives of $\mathcal{L}_{\text{int}}$ into \Eqref{eq:S_p_Gamma}, we obtain 
\begin{align}
S_{(p)} \left(\mathbf{k}\right) &= \frac{1}{\mathrm{i}} \frac{e}{4\pi^2}  \frac{m_e^2}{45} \sqrt{\frac{\rm k}{2}} \left(\frac{e}{m_e^2}\right)^3 \int \mathrm{d}^4x\; \mathrm{e}^{\mathrm{ik}(\hat{\bf k}-t)} \nonumber \\
&\phantom{=} \times \Bigl( 4\left[\mathbf{e}_{\left(p\right)}\cdot{\bf E} - \mathbf{e}_{\left(p+1\right)}\cdot \mathbf{B}\right] \mathcal{F} \nonumber\\
&\phantom{= \times \Bigl(} +7 \left[\mathbf{e}_{\left(p\right)}\cdot\mathbf{B} + \mathbf{e}_{\left(p+1\right)}\cdot {\bf E}\right] \mathcal{G} \Bigr)\,. \label{eq:S1_FG}
\end{align}

This signal amplitude can be related to the differential number of signal photons ${\rm d}^3N_{\left(p\right)}$ of polarization $p$ which have an energy ${\rm k}=|\bf{k}|$ in the differential energy interval ${\rm dk}$ and are emitted into the solid angle ${\rm d}\Omega$ 
around $\hat{{\bf k}}$ as  
\begin{equation}
{\rm d}^3N_{(p)}=\frac{{\rm k^2}{\rm dk}\,{\rm d}\Omega}{(2\pi)^3}\,|{\cal S}_{(p)}({\bf k})|^2\,.\label{eq:d3Nsig}
\end{equation}
This differential form leads to a polarization sensitive angular resolved signal photon density
\begin{equation}
\rho_{(p)}(\varphi,\vartheta|{\rm k}_{\rm min},{\rm k}_{\rm max}) = \frac{1}{(2\pi)^3} \!\int_{\mathrm{k}_{\rm min}}^{\mathrm{k}_{\rm max}} \!\!\!\mathrm{dk}\,{\rm k}^2\bigl|S_{(p)}(\mathbf{k})\bigr|^2
\label{eq:rho_result}
\end{equation}
depending on the emission angles $\varphi$ and $\vartheta$ and integrated over a given frequency range $\rm k_{\rm min}$ to $\rm k_{\rm max}$.
The total number of signal photons $N_{(p)}$ with polarization $(p)$ in a given solid angle regime $\cal A$ reads
\begin{equation}
N_{(p)}({\cal A}|{\rm k}_{\rm min},{\rm k}_{\rm max})=\int_{\cal A}{\rm d}\Omega\,\rho_{(p)}(\varphi,\vartheta|{\rm k}_{\rm min},{\rm k}_{\rm max})\,. \label{eq:NpA}
\end{equation}
If the measurement is polarization insensitive we have to sum both polarizations, i.e. $N({\cal A}|{\rm k}_{\rm min},{\rm k}_{\rm max})=\sum_{p=1}^2N_{(p)}({\cal A}|{\rm k}_{\rm min},{\rm k}_{\rm max})$.
Further details can be found in the Refs.~\cite{Karbstein:2014fva,Gies:2021ymf}. 

\section{Experimental scenario}\label{sec:exp}

We study the interaction of two strong laser pulses colliding under an angle $\vartheta_{\rm col} > 90^{\circ}$.
In this section, we introduce the experimental setup based on state-of-the-art techniques. 
Further we choose an example set of parameters for experimentally accessible quantum vacuum signals.

Here, we focus on laser pulses in the optical regime with the same photon energy of $\omega_0=1.55\,{\rm eV}$, as is available at the Advanced Titanium-Sapphire Laser (ATLAS) in the Centre for Advanced Laser Applications (CALA) facility (Garching, Germany) \cite{CALA, Hartmann:2021iyf} and the Jenaer Titan:Saphir 200 Terawatt Laser System (JETI-200) laser at Helmholtz Institut Jena (Jena, Germany) \cite{JETI-200, Doyle:2021mdt}.
These facilities operate state-of-the-art high-intensity lasers and illustrate which field strengths are already available in strong-field laser experiments.

Accordingly, we use optical laser pulses with a pulse energy of $W=25\,{\rm J}$ here.
The wavelength is $\lambda=800\,{\rm nm}$ and the pulse duration is $\tau_{\rm FWHM}=25\,{\rm fs}$ measured at full width half maximum (FWHM) of the intensity.
Assuming that the temporal shape of the pulse is Gaussian, we can convert the FHWM pulse duration $\tau_{\rm FWHM}$ into an $1/e^2$ pulse duration $\tau= 2\sqrt{\ln2}\, \tau_{\rm FWHM}$.  
The calculations are performed with the $1/e^2$ pulse duration; if in the following the FWHM duration is mentioned, then this is indicated with the addition (FWHM) in brackets.

\begin{figure}[t]
	\centering
	\includegraphics[width=\columnwidth]{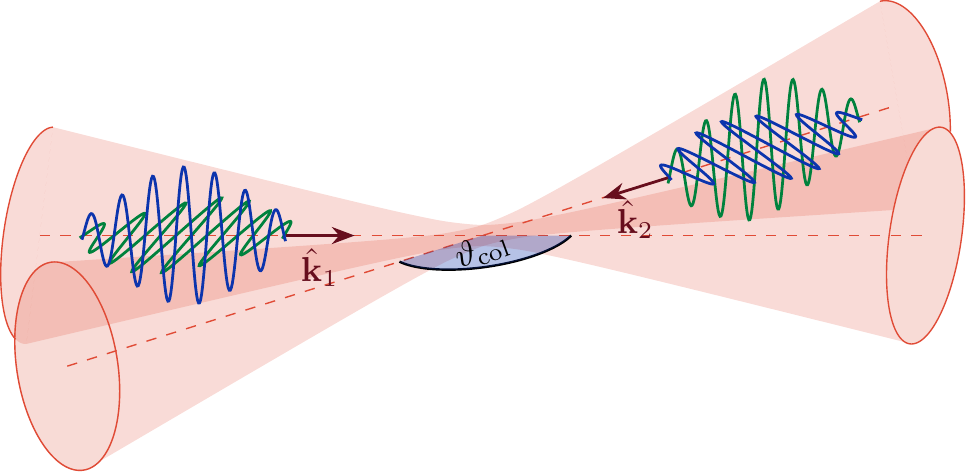}
	\caption{Illustration of the experimental scenario. The red shades illustrate the forward/backward cones of two laser pulses colliding under the angle $\vartheta_{\rm col}$ with a circular focus cross section. Additionally, the green (blue) lines represent a typical snapshot of the electric (magnetic) field amplitude.}
	\label{fig:Laser_pulses}
\end{figure}

Moreover, we model the spatial structure of the laser pulses by Gaussian beams with a focus waist $w_0$, which is determined by the radial aperture angle of the beams in the far field $\Theta$ via the relation $\Theta=\frac{w_0}{z_R}$ with the Rayleigh range $z_R=\pi\frac{w_0^2}{\lambda}$.
According to the laws of optics, the minimum focus width is $w_0\simeq\lambda$, which corresponds to an radial aperture of $\Theta=\frac{1}{\pi}\approx 18.24^{\circ}$.
Because of the scaling $\Theta\sim\frac{1}{w_0}$, setups with a focal width larger than the minimum yield smaller aperture angles in the far field. Consequently, the photon distribution of the driving lasers constituting a large background is less wide.

In this work, we consider the collision of two Gaussian laser pulses under the angle $\vartheta_{\rm col}$. For our calculations, we vary this angle in the range from $100^{\circ}$ to $160^{\circ}$ in $10^{\circ}$ steps.
For convenience, we refer to one laser as the probe field, labeled by subscript $1$, and the other as the pump field, labeled by the subscript $2$.
We deliberately avoid head-on collisions which would maximize the signal photon yield for both practical and theoretical reasons; see below.
Head-on collisions are very challenging and may not be practical due to experimental constraints.
We will focus mainly on the signal emitted in the vicinity of the forward direction of the probe beam. 
The purpose of the pump beam is to provide a localized strong field region inducing the signal. 
The propagation directions of the laser pulses are $\hat{\bf k}_1=\hat{\bf e}_z$ and $\hat{\bf k}_2=\sin\vartheta_{\rm col}\,\hat{\bf e}_x + \cos\vartheta_{\rm col}\, \hat{\bf e}_z$, respectively. 
Their wave vectors are ${\bf k}_1=\omega_1 \hat{\bf k}_1$ and ${\bf k}_2=\omega_2 \hat{\bf k}_2$ with $\omega_1=\omega_2=\omega_0=2\frac{2\pi}{\lambda}$.
These laser pulses collide at their focal spots at ${\bf x}=0$ and reach their peak fields at $t=0$, which defines the origin of our coordinate system.

The laser pulses are all linearly polarized. Here, we choose the polarization of the probe laser such that the electric field $\mathbf{E}_1({\bf x},t)$ points into $x$ direction; correspondingly the magnetic field $\mathbf{B}_1({\bf x},t)$ is orientated along the $y$ axis. 
To allow for a generic linearly polarized pump laser, we introduce the angle $\beta_2$ between $\mathbf{E}_1({\bf 0},0)$ and $\mathbf{E}_2({\bf 0},0)$ for the collinear collision with $\vartheta_{\rm col}=0^{\circ}$. 
Using $\vartheta_{\rm col}$ as the  general collision angle, we obtain
\begin{align}
\mathbf{E}_2({\bf x},t)&= \mathcal{E}_2({\bf x},t) \left(\begin{array}{c}
                                                    \cos\beta_2\,\cos\vartheta_{\rm col} \\ \sin\beta_2 \\ -\cos\beta_2\,\sin\vartheta_{\rm col}
                                                   \end{array}\right)\,,
\end{align}
$\mathbf{B}_2({\bf x},t)\sim \hat{\bf k}_2 \times \mathbf{E}_2({\bf x},t)$
for the pump fields, where $\mathcal{E}_i({\bf x},t)$ describes the general amplitude function.
The maximum of the total polarization insensitive signal occurs for an angle $\beta_2=90^{\circ}$ which is our  angle of choice unless otherwise stated.   
In a special scenario, we consider the signal of birefringence maximized at an angle of $\beta_2=45^{\circ}$, see \cite{Karbstein:2021hwc}.
Further references discuss vacuum birefringence from a general point of view, cf. \cite{DiPiazza:2006pr,Heinzl:2006xc,Karbstein:2015xra, Mosman:2021vua, Ataman:2018ucl}.

\begin{figure}[t]
	\centering
	\includegraphics[width=\columnwidth]{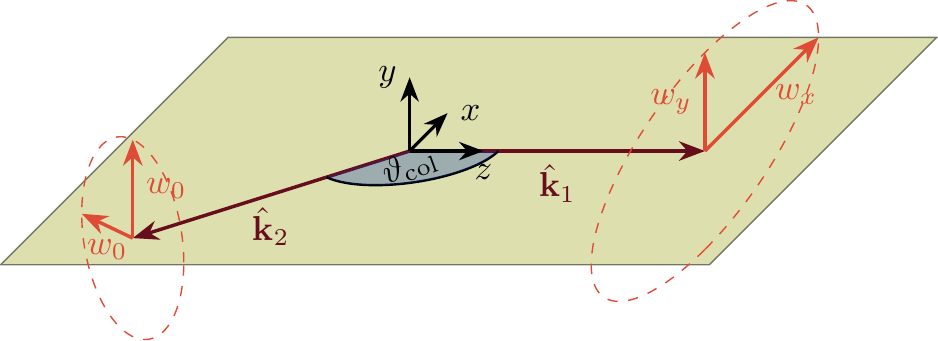}
	\caption{Illustration of the beam radii in the focus of the experimental scenario. Besides the wave vectors of probe $\hat{\rm k}_1$ and pump $\hat{\rm k}_2$, the red dashed curves illustrate the radii of the waist in the focus. The rotationally symmetric pump has a waist size of $w_0$ while the probe with elliptical cross section has the waists $w_x$ in the collision plane marked by the yellow plain and $w_y$ perpendicular to it.}
	\label{fig:Laser_pulses_Focus}
\end{figure}

Throughout this work we consider a radially symmetric pump, but use both elliptical and circular focus cross sections for the probe laser.
To this end, we introduce two independent beam waists for the probe laser $w_x$ in the collision plane and $w_y$ normal to this plane.
For convenience, we parameterize $w_x=w_y=\mu\,w_0$ in the case of a rotationally symmetric probe beam, using a
dimensionless parameter $\mu \geq 1$.
In the case of elliptical cross sections, we analogously introduce $w_x=\mu_x\,w_0$ and $w_y=\mu_y\,w_0$ with $\{\mu_x,\mu_y\}\geq 1$.
Fig. \ref{fig:Laser_pulses} illustrates this scenario with rotationally symmetric pump and probe beams.

To determine the rotationally symmetric quantum vacuum signal, it is sufficient to consider the Gaussian beams in the limit of infinite Rayleigh ranges.
This approximation is justified if the dimensions of the interaction volume are small compared to the Rayleigh lengths $z_R$.
This is true in the case of very small pulse durations $\tau_i\ll z_{R,i}$ compared to the Rayleigh ranges for all involved beams.
Another way to justify the approximation is a small ratio of focal width $w_{0,i}$ and Rayleigh length $z_{R,j}$ of two colliding beams $i$ and $j$ with respect to the sine of their collision angle, i.e. $w_{0,i}/z_{R,j}\ll \left|\sin\vartheta_{\rm col}\right|$ for $i\neq j$.
If any of the above conditions is satisfied, then the effectively interacting regions of the laser beams can be approximated as tubes in the interaction volume \cite{Gies:2017ygp, King:2018wtn, Yariv75book, Siegman86}.
Figure \ref{fig:Laser_pulses_Focus} shows the beam radii in the focus by using a probe with elliptical cross section.
For numerical approaches beyond infinite Rayleigh range approximation see \cite{Grismayer:2016cqp, Blinne:2018uyn, Blinne:2018nbd, Lindner:2021krv, Sainte-Marie:2022efx}.  

Correspondingly, the field amplitudes can be approximated as
\begin{equation}
\mathcal{E}_1 \left(x\right) = \frac{\mathcal{E}_{0}}{\sqrt{\mu_{x}\mu_{y}}}\, \mathrm{e}^{-(\frac{{\bf x}\cdot\hat{\bf k}_1-t}{\tau_1/2})^2 - \frac{x^2}{w_{x}^2} - \frac{{y}^2}{w_{y}^2} }   \cos\bigl(\omega_0({\bf x}\cdot\hat{\bf k}_1-t)\bigr)  \label{eq:E1_x_infRay}
\end{equation}
and
\begin{equation}
\mathcal{E}_2 \left(x\right) = {\mathcal{E}_{0}}\, \mathrm{e}^{-(\frac{{\bf x}\cdot\hat{\bf k}_2-t}{\tau_2/2})^2 - \frac{{\bf x}^2-\left(\bf x\cdot \hat{k}_2\right)^2}{w_{0}^2}  }   \cos\bigl(\omega_0({\bf x}\cdot\hat{\bf k}_2-t)\bigr)\,,  \label{eq:E2_x_infRay}
\end{equation}
where we have used the different propagation directions $\hat{\bf{k}}_1$ and $\hat{\bf{k}}_2$. 
Here, $\mathcal{E}_{0}$ is the field amplitude maximum depending on the laser pulse properties. 
It fulfills \cite{Karbstein:2017jgh}
\begin{equation}
\mathcal{E}_{0} = \sqrt{8\sqrt{\frac{2}{\pi}} \frac{W}{\pi w_0^2 \tau}} \label{eq:E_star} \approx 8.5\times10^{15}\,{\rm V/m}\,,
\end{equation}
using the parameters given above.

\section{Calculation of the signature of quantum vacuum nonlinearites}

\subsection{The background of the driving laser pulses}

Before we calculate the signal, let us determine the distribution of the driving laser photons far outside the interaction volume.

Since we need extremely strong laser fields to stimulate QED vacuum nonlinearites, the challenge in experiment is to discern the small quantum vacuum signal from the enormous number of photons of the driver lasers.
In the current research, there are many ideas to suppress this background such as an elastic scattering signal outside the forward cones of the driving lasers, \cite{Klar:2020ych}. 
Another idea is to use inelastic scattering to obtain signal at a frequency different from the background and thus achieve a spectral signal-to-background separation \cite{Gies:2021ymf}. 
Furthermore, in experiments on birefringence, the different polarization of signal and background can be the key to measurability \cite{Heinzl:2006xc}.

Let us describe the background quantitatively, in order to assess where the signal dominates.
In contrast to the signal, the background in the far field cannot be reliably determined in the limit of infinite Rayleigh length.

We can estimate the total number of photons per laser pulse as $N^{\rm tot}=W/\omega_0\approx 10^{20}$.
This results in a differential number of background photons of the $i$th laser ${\rm d}N^{\rm Bg}_i$ per solid angle element ${\rm d}\Omega$ which satisfies the relation $N^{\rm tot}=\int_{\Omega}\! {\rm d}N^{\rm Bg}_i$, since we use $W_1=W_2=W$ and frequency $\omega_1=\omega_2=\omega_0$ for both lasers.
For the far-field distribution of a Gaussian beam with normalization $N^{\rm Bg}_0=\frac{w_0^2\omega_0}{2\pi}W$, we obtain
\begin{equation}
\frac{{\rm d}N^{\rm Bg}_i}{{\rm d}\Omega} = \mu_{1i}\mu_{2i}\, N^{\rm Bg}_0\, {\rm e}^{ - \frac{1}{2} \omega_0^2 w_0^2\left(\mu_{1i}^2 \cos^2\!\phi_i\; + \mu_{2i}^2  \sin^2\!\phi_i\right) \theta_i^2 } \label{eq:NiBg}
\end{equation}
for a laser pulse with elliptical cross section. Here, 
\begin{equation}
 \phi_1 = \varphi
\end{equation}
and
\begin{equation}
 \phi_2 = \arctan\left( \cos\varphi\, \sin\vartheta\, \cos\vartheta_{\rm col} -  \sin\varphi\, \sin\vartheta\, \sin\vartheta_{\rm col}  \right)
\end{equation}
parameterize the rotations around the beam axis,
and 
\begin{eqnarray}
  \theta_1&=& \vartheta - \arccos\left(\hat{\rm k}_1\cdot \hat{\rm e}_z\right),\\
  \theta_2&=& \arccos\left( \cos\varphi\, \sin\vartheta\, \sin\vartheta_{\rm col} +  \cos\vartheta\, \cos\vartheta_{\rm col}  \right)
\end{eqnarray}
are the polar angles measured from the forward beam axis; $\varphi$ and $\vartheta$ are azimuthal and polar angle used to parameterize the emission direction in a spherical coordinate system with the north pole along the outgoing direction of the probe laser $\mathbf{k}_1$.

For the pump laser with a circular cross section in the far field, we have
\begin{equation}
\frac{{\rm d}N^{\rm Bg}_2}{{\rm d}\Omega} = N^{\rm Bg}_0\, {\rm e}^{ - \frac{1}{2} \omega_0^2 w_0^2 \theta_2^2 }\,.
\end{equation}
Equation \eqref{eq:NiBg} implies that the density of background photons decreases  more rapidly with $\theta_i$ the larger $\mu_{1i}$ and $\mu_{2i}$.
In the present scenario, the full density of the background photons is given by 
\begin{equation}
 \rho_{\bar{\mu}}^{\rm Bg}(\varphi, \vartheta) = \sum_{i=1}^2 \frac{{\rm d}N^{\rm Bg}_i}{{\rm d}\Omega}\,.
\end{equation}
Since we keep the total pulse energy $W$, the frequency $\omega_0$, and the pulse duration $\tau$ constant, we write $\rho_{\bar{\mu}}^{\rm Bg}(\varphi, \vartheta)$ with the tuple $\bar{\mu}=\{\mu_x,\mu_y\}$ indicating the choice of foci of the probe beam.
In the case of circularly focused pulses, we use $\bar{\mu} \rightarrow \mu$, i.e. $\mu_x=\mu_y=\mu$.

The spectral distribution of the driving laser pulses is centered around $\omega_0$, therefore the background is also dominated by the frequency $\omega_0$.

Finally, the number of background photons in a given solid angle regime $\mathcal{A}$, analogous to Eq. \eqref{eq:NpA}, is given by
\begin{equation}
N_{\rm Bg}(\mathcal{A},\bar{\mu})=\int_{\cal A}{\rm d}\Omega\,\rho_{\bar{\mu}}^{\rm Bg}(\varphi,\vartheta)\,. \label{eq:NBg}
\end{equation}
\subsection{Quantum vacuum signal} \label{sec:sqn}

For the signal amplitude $S_{(p)}\left({\bf k}\right)$, let us
express the wave vector in spherical coordinates as ${\bf k} = {\rm k}\, \hat{\bf k}$ with $\hat{\bf k} = \cos\varphi\, \sin\vartheta\, \hat{\bf e}_x + \sin\varphi\, \sin\vartheta\, \hat{\bf e}_y +\cos\vartheta\, \hat{\bf e}_z$.

The vectors orthogonal to $\hat{\bf k}$ can then be parameterized by a single angle $\beta$ as
\begin{equation}
{\bf e}_{\beta}=\sin \beta \left.\hat{\bf k}\right|_{\varphi\rightarrow\varphi+\frac{\pi}{2},\vartheta=\frac{\pi}{2}} + \cos \beta \left.\hat{\bf k}\right|_{\vartheta\rightarrow\vartheta+\frac{\pi}{2}} \,.
\end{equation}
Without loss of generality we associate the polarization $p=1$ with $\beta$ and $p=2$ with $\beta+\frac{\pi}{2}$, respectively.

The leading quantum vacuum signal arises from the effective coupling of three electromagnetic fields.
For convenience, we express the electric and the magnetic field of the $i$th laser pulse field strength and direction according to ${\bf E}_i({\bf x},t)=\mathcal{E}_i({\bf x},t)  \hat{\bf E}_i$ and ${\bf B}_i({\bf x},t)=\mathcal{E}_i({\bf x},t) \hat{\bf B}_i$, respectively.
In turn, Eq. \eqref{eq:S1_FG} can be written as \cite{Gies:2021ymf} 
\begin{align}
 S_{(p)} \left(\mathbf{k}\right) &= \frac{1}{\mathrm{i}} \frac{e}{4\pi^2}  \frac{m_e^2}{45} \sqrt{\frac{\rm k}{2}} \left(\frac{e}{m_e^2}\right)^3 \left(\mathcal{I}_{121}g_{121} + \mathcal{I}_{212}g_{212}\right) \label{eq:S_p_Ig}
\end{align}
with
\begin{align}
 g_{iji}(\varphi,\vartheta) &= 2 \left({\bf e}_{\beta}\cdot\hat{\bf E}_i - {\bf e}_{\beta+\frac{\pi}{2}}\cdot\hat{\bf B}_i \right) \nonumber \\
 &\phantom{= - } \times \left( \hat{\bf B}_i\cdot \hat{\bf B}_j - \hat{\bf E}_i \cdot \hat{\bf E}_j \right) \nonumber \\
 &\phantom{=} - \frac{7}{2} \left({\bf e}_{\beta}\cdot\hat{\bf B}_i + {\bf e}_{\beta+\frac{\pi}{2}}\cdot\hat{\bf E}_i \right) \nonumber \\
 &\phantom{= - } \times \left( \hat{\bf B}_i\cdot \hat{\bf E}_j + \hat{\bf B}_j \cdot \hat{\bf E}_i \right)\,, \label{eq:g_func}
\end{align}
and  
\begin{align}
\mathcal{I}_{iji}\left({\bf k}\right) = \int \mathrm{d}^4x\; \mathrm{e}^{\mathrm{ik}(\hat{\bf k}-t)} \mathcal{E}_i({\bf x},t) \mathcal{E}_j({\bf x},t) \mathcal{E}_i({\bf x},t)\,.
\end{align}
The function $g_{ijl}(\varphi,\vartheta)$ depends on the orientation of the electromagnetic fields of the driving laser fields and the signal; $I_{ijl}\left({\bf k}\right)$ is the Fourier transform of the product of three field amplitudes evaluated on shell.

In the conventions used here, the relevant contributions read
\begin{align}
 g_{121}(\varphi,\vartheta)& = \frac{1}{4} \left(1 - \cos\vartheta\right) \left(1-\cos\vartheta_{\rm col}\right) \nonumber \\
 & \phantom{=} \times \left[3 \cos \left(\beta + \beta_2 -\varphi\right) - 11 \cos\left(\beta - \beta_2 -\varphi\right) \right]
\end{align}
and
\begin{align}
 g&_{212}(\varphi,\vartheta) \nonumber \\
 &= \frac{1}{4} \left(1 - \cos\vartheta_{\rm col}\right) \Big( [11\cos\beta-3\cos\left(\beta+2\beta_2\right)]  \nonumber \\
 & \phantom{=}  \times \left[\left(\cos\vartheta\,\cos\vartheta_{\rm col} -1\right) \cos\varphi + \sin\vartheta\, \sin\vartheta_{\rm col} \right] \nonumber \\
 & \phantom{=} + \left[11 \sin\beta -3 \sin \left(\beta + 2 \beta_2\right)  \right] \left(\cos\vartheta-\cos\vartheta_{\rm col}\right)\sin\varphi\Big)\,.
\end{align}

We aim at the density of signal photons $\rho_{\bar{\mu}}^{\rm Sig}(\varphi,\vartheta)$ regardless of their frequency. The latter lies within a small frequency range around $\omega_0$, see also the detailed analysis in \cite{Gies:2021ymf}. Apart from Sect. \ref{sec:qBi}, we consider the inclusive average over polarizations.
We are mainly interested in the parametric dependence of the signal density on the focus width (along both orthogonal axes) explicitly indicated by the parameter tuple $\bar{\mu}$.

We begin by substituting the signal amplitude from Eq. \eqref{eq:S_p_Ig} into Eq. \eqref{eq:rho_result}, 
\begin{align}
 \rho_{\bar{\mu}}^{\rm Sig}(\varphi,\vartheta) &= \sum_{p=1}^2 \frac{1}{(2\pi)^3} \!\int_0^{\infty} \!\!\!\mathrm{dk}\,{\rm k}^2\bigl|S_{(p)}(\mathbf{k})\bigr|^2  \nonumber \\ 
 &= \frac{1}{2\left(2\pi\right)^7} \frac{1}{45^2} \left(\frac{e}{m_e}\right)^8 \nonumber \\
 &\phantom{=}\times \!\int_0^{\infty} \!\!\!\mathrm{dk}\,{\rm k}^3   \sum_{p=1}^2 \left(\mathcal{I}_{121}g_{121}+\mathcal{I}_{212}g_{212}\right)^2 \label{eq:rhoSigIg},
\end{align}
taking advantage of the fact that the Fourier transform of a real symmetric function is also a real symmetric function, i.e., $\mathcal{I}^*_{ijl}=\mathcal{I}_{ijl}$.
All contributions $\mathcal{I}_{121}g_{121}$ and $\mathcal{I}_{212}g_{212}$ give rise to non-vanishing contributions for complementary values of $\left(\varphi,\vartheta\right)$.
The interference term $2\mathcal{I}_{121}\mathcal{I}_{212}g_{121}g_{212}$ is exponentially suppressed in comparison to the ``direct'' channels and can be safely neglected.
Therefore, we neglect the interference term in the remainder.
This also increases the performance of the numerical integration over the solid angle.

With these approximations, we can calculate the Fourier integrals $\mathcal{I}_{iji}$ analytically, since their evaluation amounts to performing Gaussian integrals. 
Then, Eq. \eqref{eq:rhoSigIg} can be expressed as
\begin{align}
 \rho_{\bar{\mu}}^{\rm Sig}(\varphi,\vartheta) &=\frac{1}{\left(2\pi\right)^7} \frac{2}{45^2} \left(\frac{e}{m_e}\right)^8 \nonumber \\
 &\phantom{=}\times \Big[ \left(\sum_{p=1}^2 g_{121}^2(\varphi,\vartheta)\right) \!\int_{0}^{\infty} \!\!\!\mathrm{dk}\,{\rm k}^3\, \mathcal{I}_{121}^2({\bf k})   \nonumber \\
  &\phantom{=\times} +  \left(\sum_{p=1}^2 g_{212}^2(\varphi,\vartheta)\right) \!\int_{0}^{\infty} \!\!\!\mathrm{dk}\,{\rm k}^3\, \mathcal{I}_{212}^2({\bf k}) \Big] \,. \label{eq:rhoSigmubar}
\end{align}

Subsequently we are mainly interested in the case of $\beta_2=\frac{\pi}{2}$, where we obtain 
\begin{align}
\sum_{p=1}^2\left.g_{121}^2(\varphi,\vartheta)\right|_{\beta_2=\frac{\pi}{2}} &=  196\, \sin^4\frac{\vartheta}{2}\, \sin^4\frac{\vartheta_{\rm col}}{2},\\
\sum_{p=1}^2\left.g_{212}^2(\varphi,\vartheta)\right|_{\beta_2=\frac{\pi}{2}} &=  49\, \sin^4\frac{\vartheta_{\rm col}}{2} \Big[\cos\vartheta\cos\vartheta_{\rm col} -1 \nonumber \\
&\qquad\qquad\qquad + \cos\varphi\sin\vartheta\sin\vartheta_{\rm col} \Big]^2\,.
\end{align}

So far, all calculations could be performed analytically.
The number of signal photons $N_{\rm Sig}(\mathcal{A},\bar{\mu})$ emitted into a given solid angle area $\mathcal{A}$ is finally evaluated by a numerical integration over $\mathcal{A}$ as
\begin{equation}
N_{\rm Sig}(\mathcal{A},\bar{\mu})=\int_{\cal A}{\rm d}\Omega\,\rho_{\bar{\mu}}^{\rm Sig}(\varphi,\vartheta)\,. \label{eq:Nsig}
\end{equation}

\section{Discernibly analysis of quantum vacuum signal} \label{sec:anal}

Our goal is to distinguish the weak quantum vacuum signal from the strong laser background.
This requires a comparison of angular resolved densities or particle numbers of the signal and background.
In this section, we discuss ways to achieve optimal signal-to-background separation.

Let is first introduce a discernibility criterion.  

We call a signal \textit{discernible} if the density of signal photons $\rho^{\rm Sig}_{\bar{\mu}}(\varphi,\vartheta) $ in a given solid-angle region $\mathcal{A}_{\bar{\mu}}$ is larger than the density $\rho^{\rm Bg}_{\bar{\mu}}(\varphi,\vartheta) $ of the background provided by the driver lasers  \cite{Karbstein:2019dxo}.
The union of all such regions defines the solid-angle region 
\begin{equation}
 \mathcal{A}_{\rm d,\bar{\mu}} = \left\{ (\varphi,\vartheta)\in\left[0,2\pi\right]\times\left[0,\pi\right) \left.\right| \rho^{\rm Sig}_{\bar{\mu}}(\varphi,\vartheta) \geq \rho^{\rm Bg}_{\bar{\mu}}(\varphi,\vartheta) \right\}\,.
\end{equation}
For simply connected $\mathcal{A}_{\rm d,\bar{\mu}}$, we describe the boundary $\partial\mathcal{A}_{\rm d,\bar{\mu}}$ using the upper boundary $\vartheta_{\rm u}(\varphi)$ and lower boundary $\vartheta_{\rm l}(\varphi)$, such that 
\begin{equation}
 \partial\mathcal{A}_{\rm d,\bar{\mu}} = \left\{\vartheta_{\rm l}(\varphi)\left.\right| \varphi\in\left[\varphi_{\rm i},\varphi_{\rm f}\right] \right\}
\cup\left\{\vartheta_{\rm u}(\varphi)\left.\right| \varphi\in\left[\varphi_{\rm i},\varphi_{\rm f}\right] \right\}\,,
\end{equation}
where $\vartheta_{\rm u}(\varphi_{\rm i})=\vartheta_{\rm l}(\varphi_{\rm i})$ and $\vartheta_{\rm u}(\varphi_{\rm f})=\vartheta_{\rm l}(\varphi_{\rm f})$, respectively.
In appendix \ref{sec:Ad}, a method for numerically determining region $\mathcal{A}_{\rm d,\bar{\mu}}$ including its boundary is detailed.
In addition, we also show there how to compute the boundary functions $\vartheta_{\rm l/u}(\varphi)$ for non-simply-connected regions.

\subsection{Probe with circular cross section} \label{sec:circ}

\begin{figure}[h]
	\centering
	\includegraphics[width=\columnwidth]{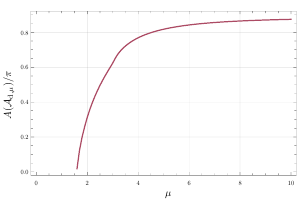}
	\caption{Area of the solid angle region $\mathcal{A}_{\rm d, \mu}$ where $\rho^{\rm Sig}_{\mu}>\rho^{\rm Bg}_{\mu}$ as a function of $\mu$ for a collision angle of $\vartheta_{\rm col}=160^{\circ}$. Here we consider the collision of two frequency $\omega_0$ beams of equal pulse duration $\tau=25\,{\rm fs}$ (FWHM) and pulse energy $W=25\,{\rm J}$. The pump (probe) beam has a circular cross section with waist radius $w_0=\frac{2\pi}{\omega_0}$ ($w_x=w_y=\mu w_0$).}
	\label{fig:Ad160}
\end{figure}

\begin{figure*}[t]
	\centering
	\includegraphics[width=\textwidth]{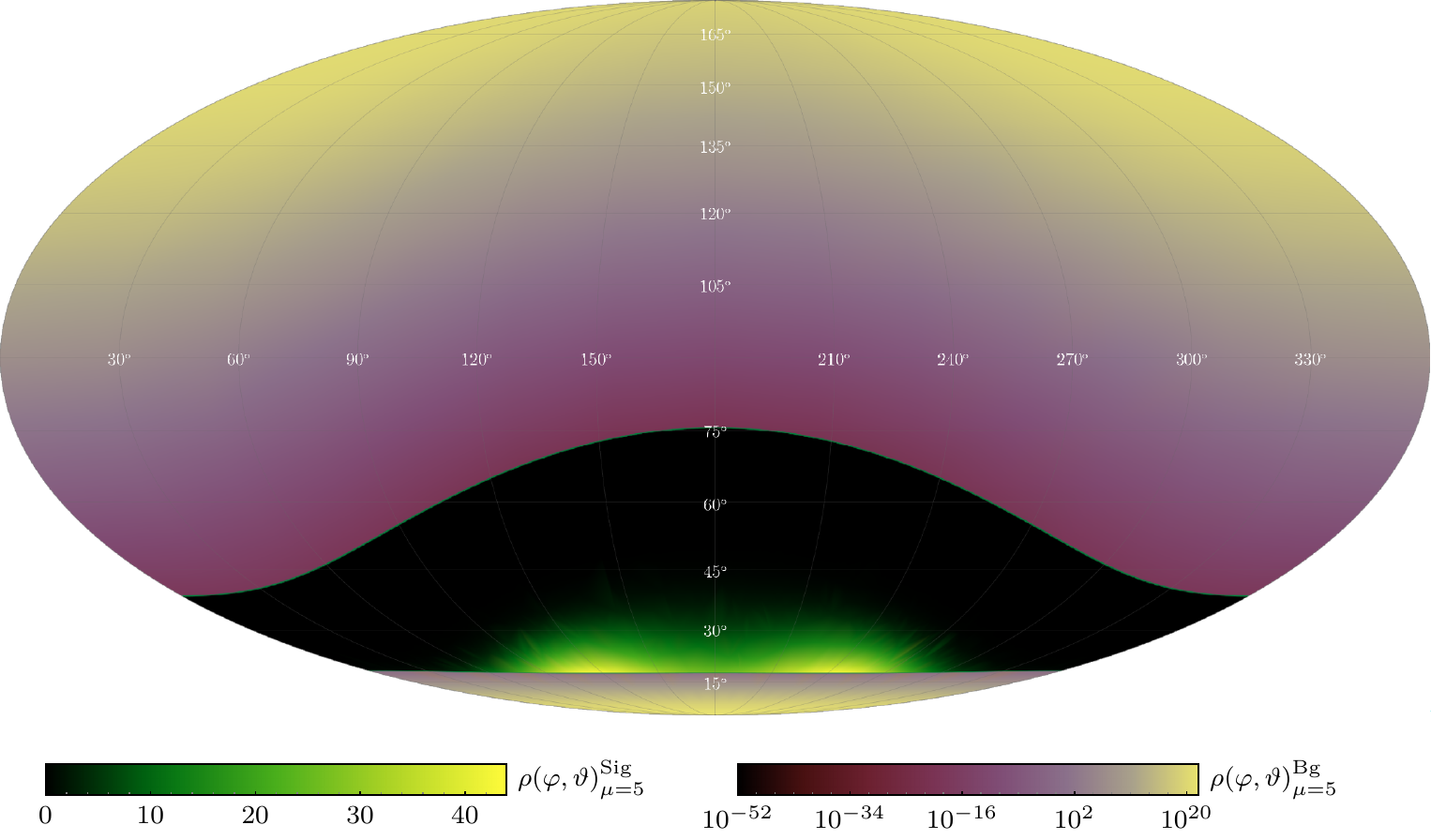}
	\caption{Mollweide plot (longitude $\varphi$, latitude $\vartheta$) of the signal photon density $\rho_{5}^{\rm Sig}(\varphi,\vartheta)$ and the background of the driving laser fields $\rho_{5}^{\rm Bg}(\varphi,\vartheta)$. The region where $\rho^{\rm Sig}>\rho^{\rm Bg}$ is highlighted by the green frame with the green color function and for the complementary region $\rho^{\rm Sig}<\rho^{\rm Bg}$ the purple color function is chosen. We consider the collision of two frequency $\omega_0$ beams of equal pulse duration $\tau=25\,{\rm fs}$ (FWHM) and pulse energy $W=25\,{\rm J}$ for $\vartheta_{\rm col}=160^{\circ}$. The pump (probe) beam has a circular cross section with waist radius $w_0=\frac{2\pi}{\omega_0}$ ($w_x=w_y=5 w_0$).}
	\label{fig:circMollSig}
\end{figure*}
Let us begin with the collision of pump and probe laser pulses with circular cross section.
First we turn our attention to the collision under the angle $\vartheta_{\rm col}=160^{\circ}$.
The pump laser is focused to its diffraction limit with $w_0=w_0=w_0=2\pi/\omega_0$, whereas the probe beam has a variable waist size $w=w_x=w_y=\mu\, w_0$ with $\mu\in[1,10]$.
The lower bound corresponds to the diffraction limit. Larger values of $\mu$ are not considered, because  a saturation is observed already before $\mu=10$. 
In the remainder of this work, we stick to the definitions and laser parameters as introduced in Sect.~\ref{sec:exp}, i.e., both pulses have a pulse duration of $\tau_1=\tau_2=25\,{\rm fs}$ (FWHM), a pulse energy of $W_1=W_2=25\,{\rm J}$ and $\omega_1=\omega_2=\omega_0=1.55\,{\rm eV}$.

At first we determine the solid angle regions where the signal is discernible; see App.~\ref{sec:Ad}. 
For convenience, we define the area of a given solid angle region $\cal A$ as 
\begin{equation}
 A(\mathcal{A}) = \int_{\cal A}{\rm d}\Omega\,.
\end{equation}
The dependence of this quantity on the parameter $\mu=w/w_0$ is shown in Fig.~\ref{fig:Ad160}.
From Fig.~\ref{fig:Ad160}, we infer that there is no discernible signal for $\mu\leq1.6$.
For $\leq\mu\lesssim4$ the area of ${\cal A}_{\rm d}$ grows very strongly, but the increase with $\mu$ becomes much slower beyond $\mu\approx4$.

In Fig.~\ref{fig:circMollSig}, we highlight the photon density for the full solid angle region of $4\pi$ for the example $\mu=5$.
Here, the region where the signal dominates the background is framed in green. 
Inside this region, the density of the signal photons is shown in the corresponding color scale (from black via green to yellow).
Outside this area, the background of the driving photons is shown; for this a different color scale (from black via purple to ocher) is used.

\begin{figure}[t]
	\centering
	\includegraphics[width=\columnwidth]{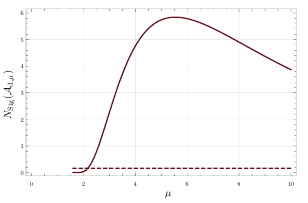}
	\caption{Number of discernible signal photons $N_{\rm Sig}(\mathcal{A}_{\rm d,\mu})$ as a function of $\mu$ for a collision angle of $\vartheta_{\rm col}=160^{\circ}$. The maximum number of background photons $N_{\rm Bg}(\mathcal{A}_{\rm d, 3.8})\approx0.16$ is reached at $\mu=3.8$. See Figure \ref{fig:Ad160} for the laser parameters employed here.}
	\label{fig:circNSig}
\end{figure}
\begin{figure*}[t]
	\centering
	$\vartheta_{\rm col}=100^{\circ}$\includegraphics[width=0.32\textwidth]{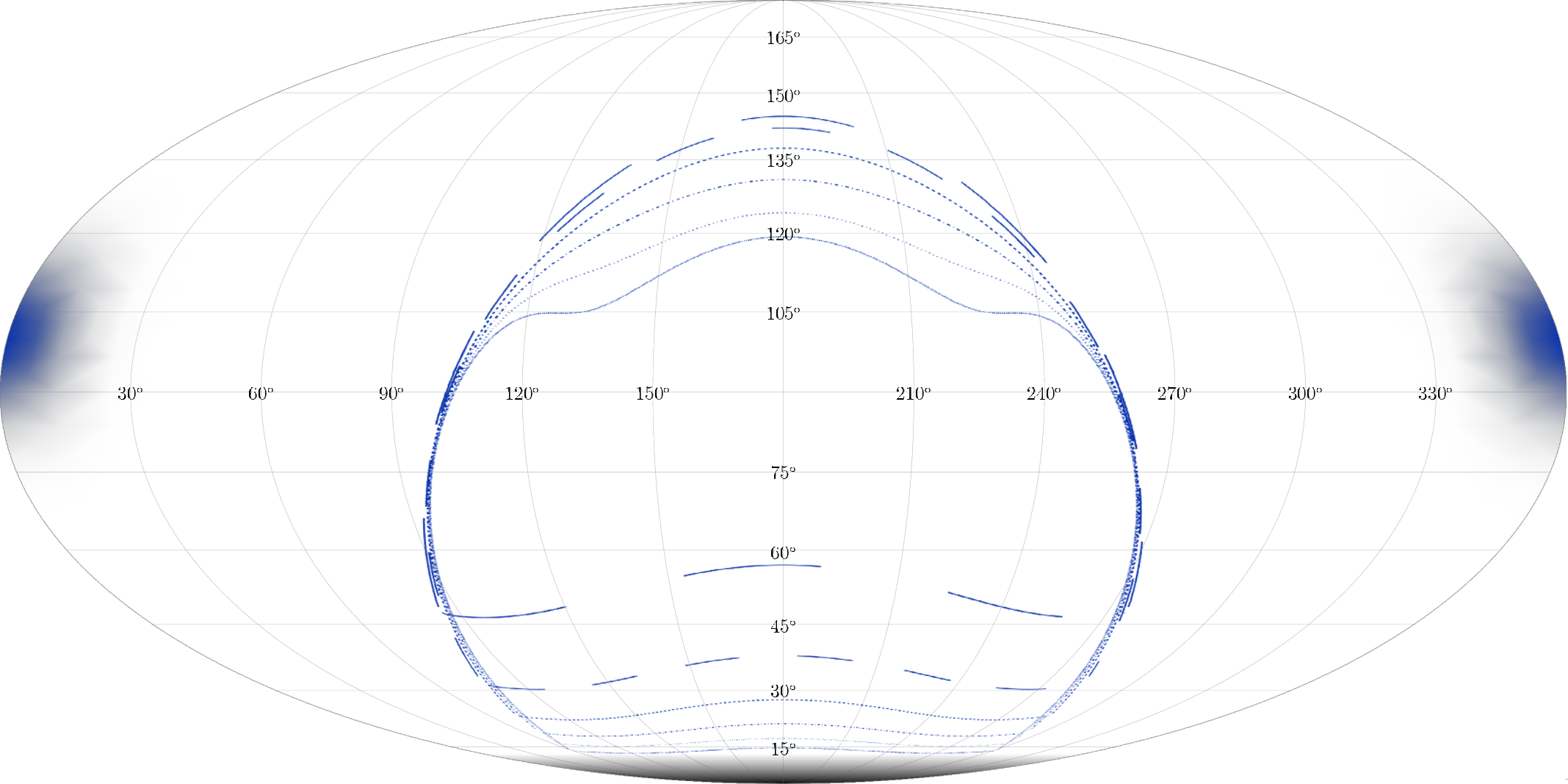}
	$\qquad\vartheta_{\rm col}=110^{\circ}$\includegraphics[width=0.32\textwidth]{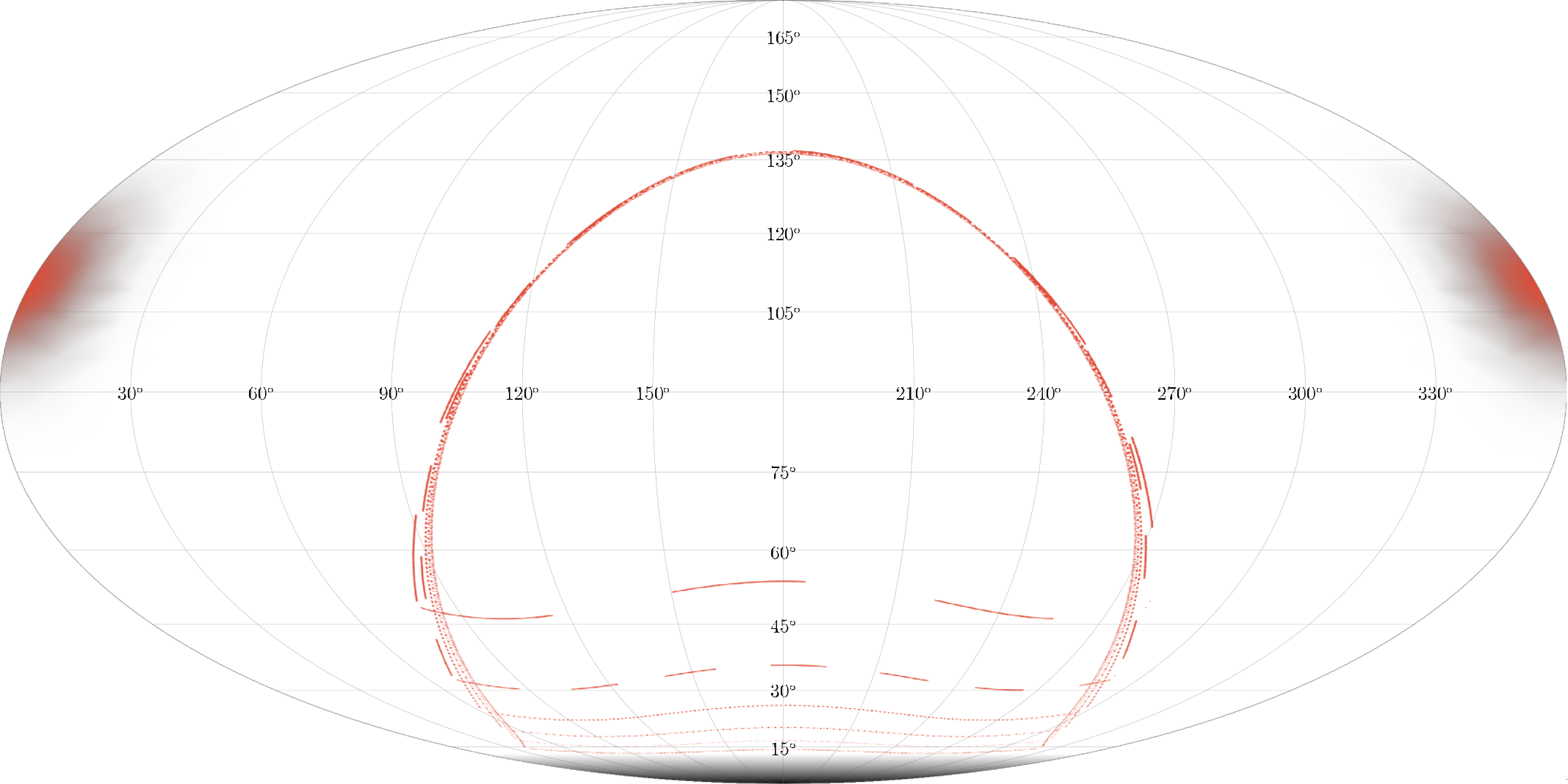}
	$\vartheta_{\rm col}=120^{\circ}$\includegraphics[width=0.32\textwidth]{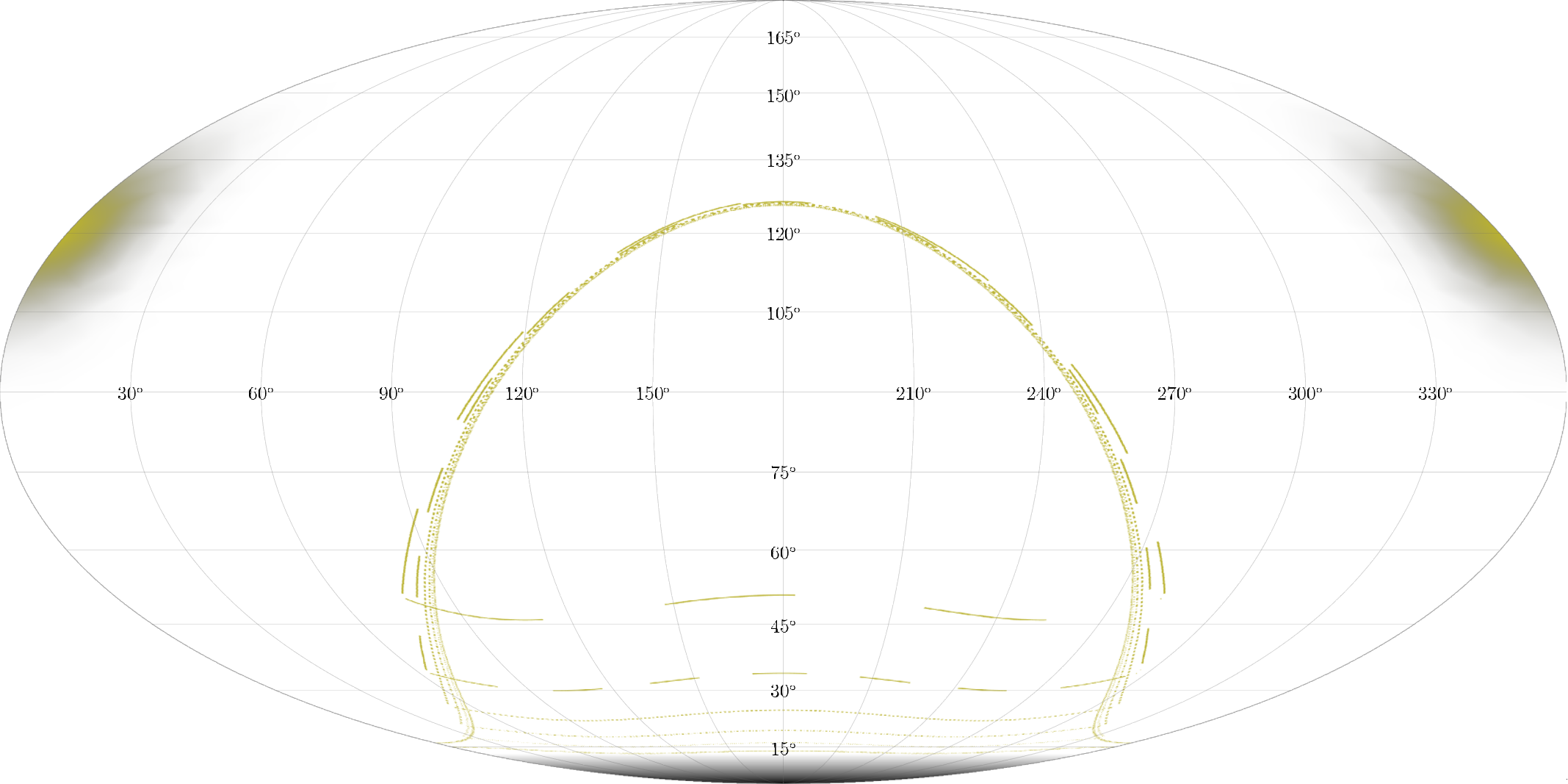}
	$\qquad\vartheta_{\rm col}=130^{\circ}$\includegraphics[width=0.32\textwidth]{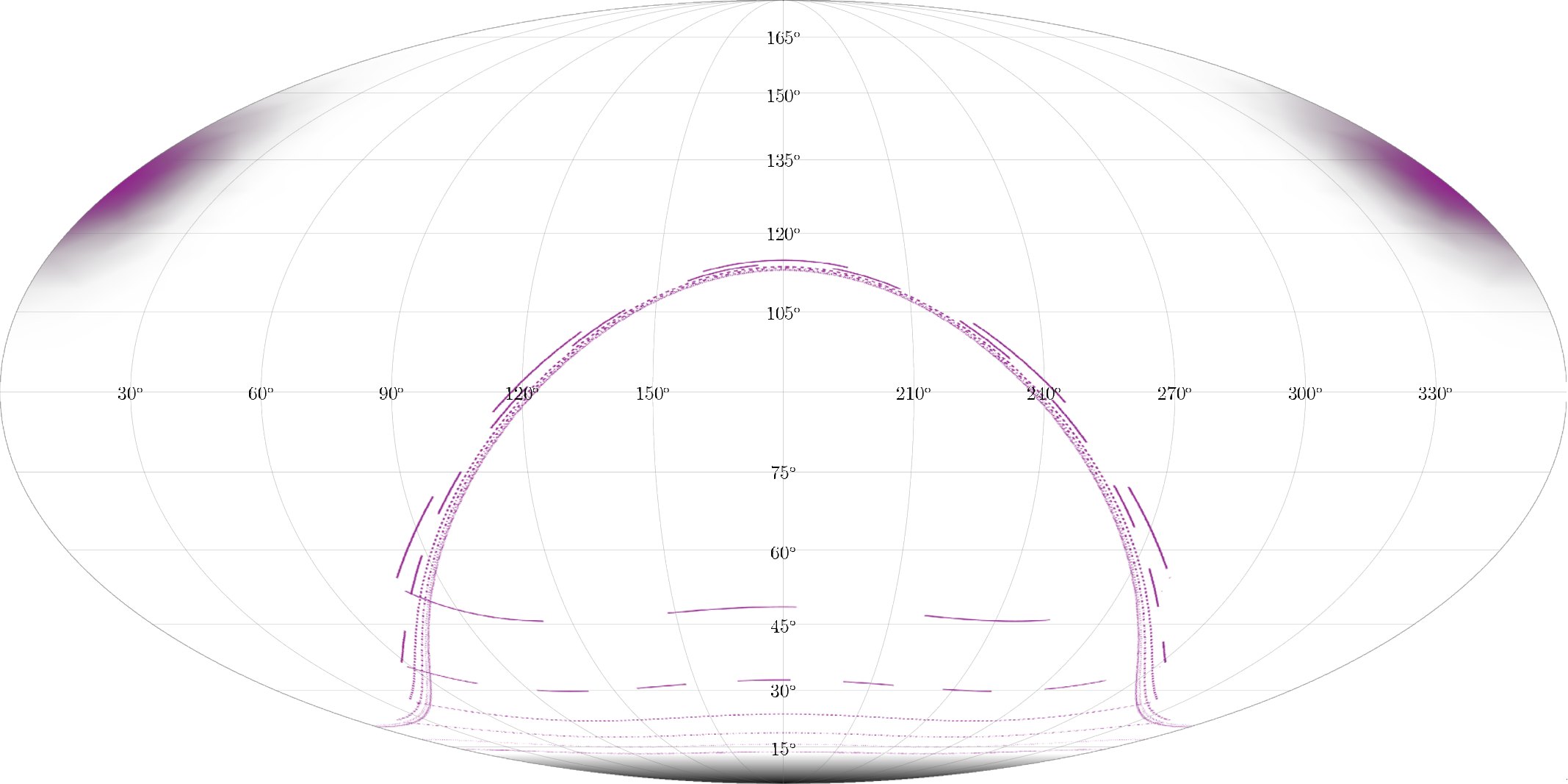}
	$\vartheta_{\rm col}=140^{\circ}$\includegraphics[width=0.32\textwidth]{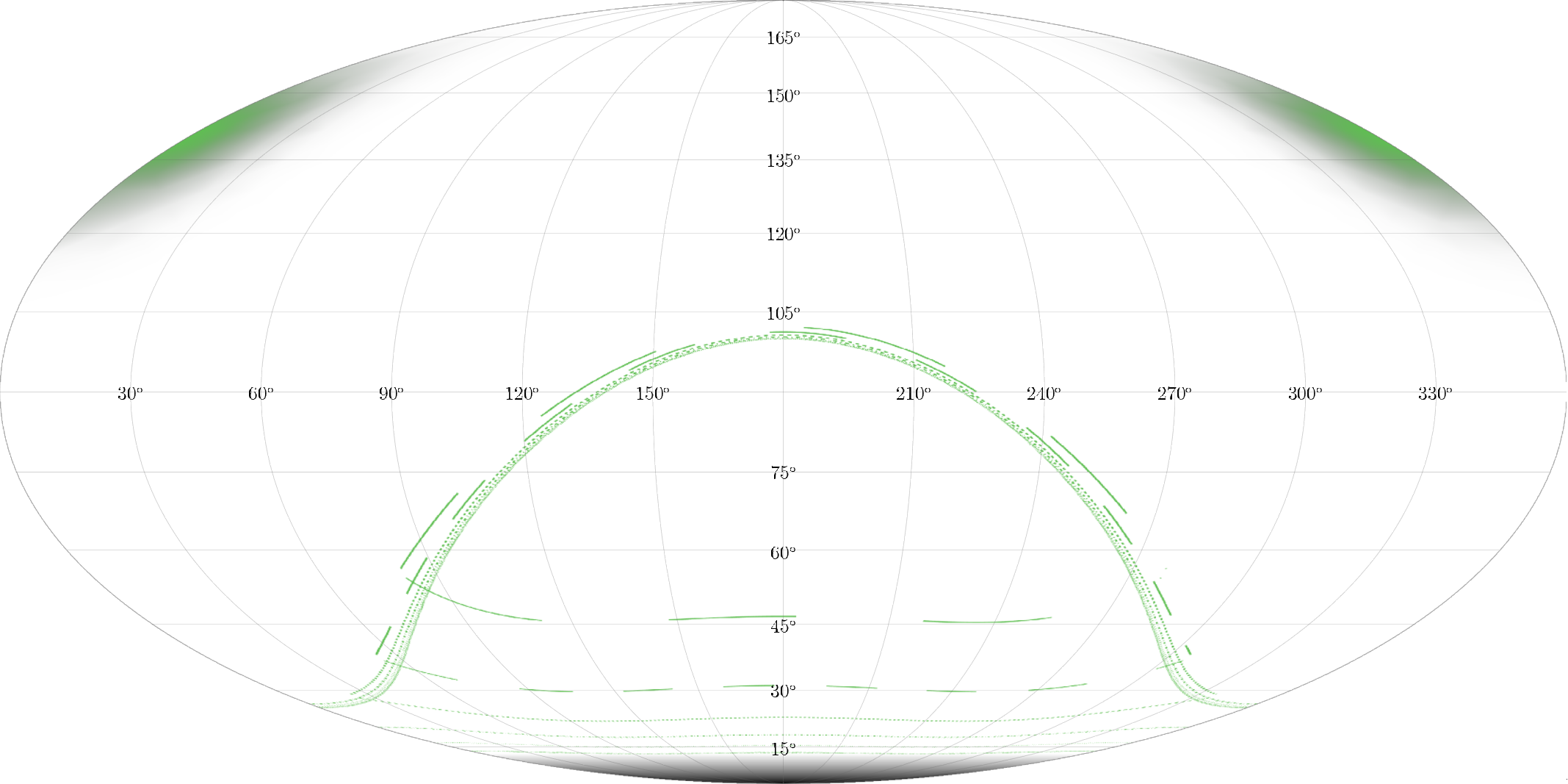}
	$\qquad\vartheta_{\rm col}=150^{\circ}$\includegraphics[width=0.32\textwidth]{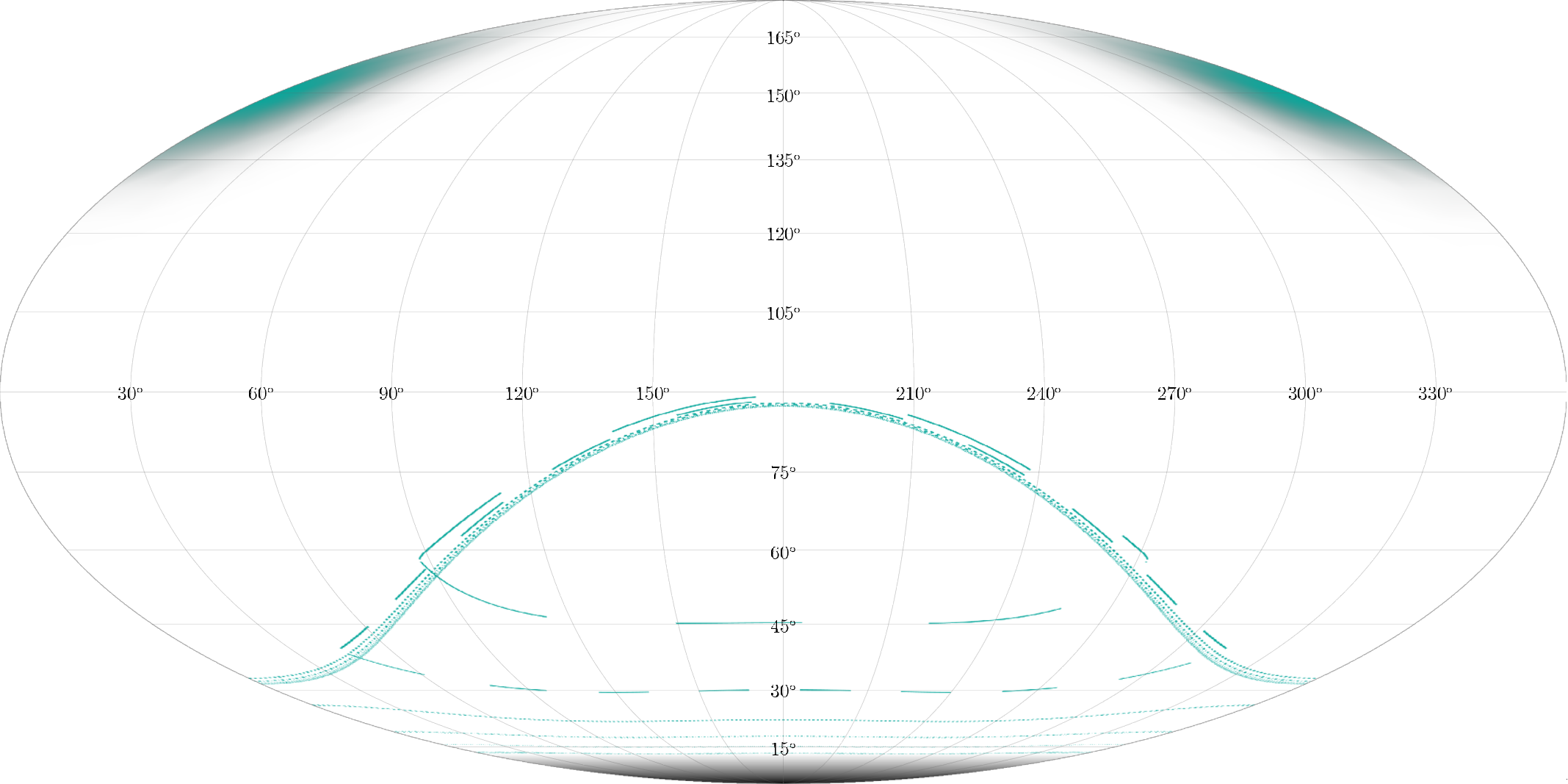}
	$\vartheta_{\rm col}=160^{\circ}$\includegraphics[width=0.32\textwidth]{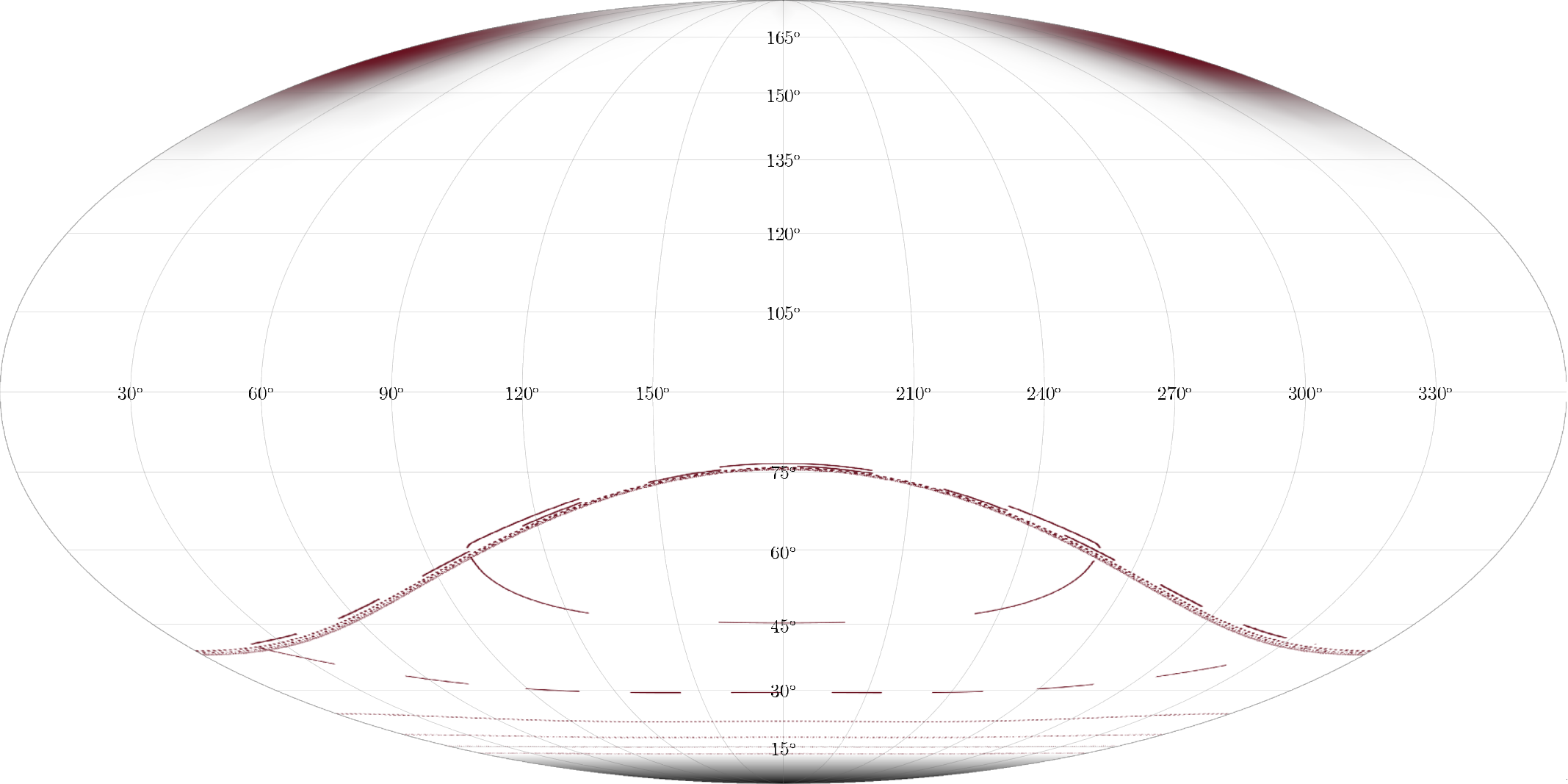}
	\caption{Mollweide plot (longitude $\varphi$, latitude $\vartheta$) showing the different direction of the pump laser pulses for different collision angles $\vartheta_{\rm col}$. The forward beam axis of the pump laser is marked by different colors and the forward beam axis of the probe is marked by the black color. The bounds $\vartheta_{\rm l}(\varphi)$ and $\vartheta_{\rm u}(\varphi)$ highlighting $\rho^{\rm Sig}=\rho^{\rm Bg}$ are marked by colored lines for different values of $\mu$: $2$, $3$, $4$, $5$, $6$, $7$. The value of $\mu$ is encoded by the lines style. Lager $\mu$ means smaller spacing between dashes. The pump (probe) is focused to a waist of $w_0=\frac{2\pi}{\omega_0}$ ($w_x=w_y=\mu\omega_0$). Both frequencies $\omega_0=1.55\,{\rm eV}$ lasers have the same pulse energy $W=25\,{\rm J}$ and duration $\tau=25\,{\rm fs}$ (FWHM).}
	\label{fig:circMollSigMultiTheta}
\end{figure*}
\begin{figure}[t]
	\centering
	\includegraphics[width=\columnwidth]{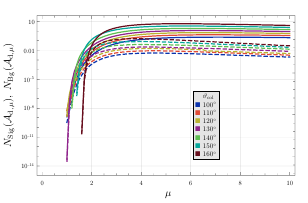}
	\caption{Number of signal (solid) and background photons (dashed) in solid angle region $\mathcal{A}_{\rm d, \mu}$ as a function of $\mu=w_x/w_0=w_y/w_0$ for different collision angles $\vartheta_{\rm col}$. See Fig. \ref{fig:circMollSigMultiTheta} for the laser parameters employed here.}
	\label{fig:circNSigMultiTheta}
\end{figure}

In order to map the sphere onto a flat diagram, we use a Mollweide projection, which
conserves the relevant areas of surfaces but is not angle conserving. 
The lower boundary $\vartheta_{\rm l}(\varphi)$ is almost constant in $\varphi$ because this region is close to the forward beam axis of the probe beam  $\vartheta=0$ which has a circular cross section.

By contrast, the upper bound $\vartheta_{\rm u}(\varphi)$ shows a more pronounced $\varphi$ dependence. 
It becomes minimal in the collision plane of the driving lasers at $\varphi=0$ and maximal
 perpendicular to the collision plane at $\varphi=180^{\circ}$.
In between,  $\vartheta_{\rm u}(\varphi)$ varies monotonically.

The color scale encodes the distribution of the signal photons. 
The largest discernible signals are reached close to $\vartheta_{\rm l}(\varphi)$.
Furthermore, we observe pronounced maxima for $\vartheta\approx 17,4^{\circ}$ and $\varphi\in[100^{\circ},140^{\circ}]$ and $\varphi\in[220^{\circ},260^{\circ}]$, respectively.

Next, we analyze the total number of discernible signal photons $N_{\rm Sig}(\mathcal{A}_{\rm d,\mu})$ as a function of $\mu$, as depicted in Fig.~\ref{fig:circNSig}. 
Only for $\mu\gtrsim1.6$, a discernible signal can be identified.
The increase of $N_{\rm Sig}$ differs visibly from the growth of $\mathcal{A_{\rm d,\mu}}$ with $\mu$.
At $\mu\approx5.5$ we encounter a pronounced maximum with $N_{\rm Sig}(\mathcal{A}_{\rm d,5.5})\approx 5.84$ photons per shot.
For comparison, the associated number of background photons yields $N_{\rm Bg}(\mathcal{A}_{\rm d,5})\approx 0.12$ photons per shot.
For $\mu\gtrsim 5.5$ $N_{\rm Sig}$ decreases approximately linearly with $\mu$.
A similar behavior was identified in the head-on collision of optical and x-ray laser pulses in \cite{Mosman:2021vua}.

The appearance of a local maximum in $N_{\rm Sig}$ shows that the probe waist provides a handle to amplify or decrease the signal for given other parameters.
Since the pulse energy and duration are kept constant, the probe intensity decreases for a wider focus. 
At the same time, a wider focus tends to increase $\mathcal{A}_{\rm d,\mu}$ because the probe divergence diminishes. 
Even though the area $A(\mathcal{A}_{\rm d, \mu})$ does not change too much, the value of the lower limit $\vartheta_{\rm l}$ changes significantly; this can be seen in Fig.~\ref{fig:circMollSigMultiTheta}.
At some point, however, the intensity of the probe eventually becomes too weak for large values of $\mu$ such that the number of signal photons decreases.

In Fig.~\ref{fig:circNSig}, apart from $N_{\rm Sig}(\mathcal{A}_{\rm d, \mu})$, the maximum number of background photons ${\rm max}\{N_{\rm Bg}(\mathcal{A}_{\rm d, \mu})|{\mu\in[1.6,10]}\}\approx 0.16$ is marked with a dashed line. 
This maximum is reached for $\mu=3.3$. 

In Fig.~\ref{fig:circMollSigMultiTheta}, we highlight our results for $\mathcal{A}_{\rm d,\mu}$ for other collision angles. 
Here we analyze the collision angles from $100^{\circ}$ to $160^{\circ}$ in $10^{\circ}$ steps.

This figure shows the boundaries $\vartheta_{\rm l}(\varphi)$ and $\vartheta_{\rm d}(\varphi)$ of the regions $\mathcal{A}_{\rm d,\mu}$.
Besides, the directions of the driving lasers are indicated: the colored shading indicates the direction of the pump laser, the black one the direction of the probe beam.

For each value of $\vartheta_{\rm col}$, we plot $\partial\mathcal{A}_{\rm d,\mu}$ for six different values of $\mu\in\{2,3,4,5,6,7\}$. 
These are indicated by the spacing of the dashing; the larger $\mu$, the smaller the spacing of the dashes.

We observe that the upper bound $\vartheta_{\rm u}$ is almost independent of $\mu$ for $\vartheta_{\rm col}>100^{\circ}$.
The reason for this is that the choice of $\mu$ affects only the focusing of the probe laser propagating along $\vartheta=0$. 
At the same time, we find that the lower bound $\vartheta_{\rm l}$ becomes smaller, i.e., $A(\mathcal{A}_{\rm d,mu})$ grows with increasing $\mu$.

Analogous to Fig.~\ref{fig:circNSig} for $\vartheta_{\rm col}=160^{\circ}$, we plot the number of signal and background photons per shot in the regions $\mathcal{A}_{\rm d,\mu}$ as a function of $\mu$ in Fig.~\ref{fig:circNSigMultiTheta}.
Here the different photon numbers are compared adapting a logarithmic scale.
In each case, the number of signal photons (solid) surpasses the number of background photons (dashed) in line with the discernibility criterion adopted to determine $\mathcal{A}_{\rm d,\mu}$. 
For each value of $\vartheta_{\rm col}$ considered here, a local maximum in the numbers of signal or background photons is observed.
The positions $\mu_{\rm max}$ of these maxima shift to smaller values of $\mu$ as the collision angle $\vartheta_{\rm col}$ increases. 
Moreover, we find that the position of the maximum of the background varies slower with $\mu$ than that for the signal when changing $\vartheta_{\rm col}$.
In addition, $\mu^{\rm Bg}_{\rm max}<\mu^{\rm Sig}_{\rm max}$ always holds in the considered cases.

\begin{table}
\begin{tabular}{|*{4}{*{1}{|c}|}|}
$\vartheta_{\rm col}$ & $\mu=\mu_{\rm max}^{\rm Sig}$ & $N_{\rm Sig}({\cal A}_{{\rm d},\mu})$ &  $N_{\rm Bg}({\cal A}_{{\rm d},\mu})$ \\ \hline
$100^{\circ}$ & $7.1$ & $0.237$ & $0.004$ \\
$110^{\circ}$ & $6.9$ & $0.414$ & $0.006$ \\
$120^{\circ}$ & $6.5$ & $0.694$ & $0.012$ \\
$130^{\circ}$ & $6.4$ & $1.139$ & $0.017$ \\
$140^{\circ}$ & $6.1$ & $1.898$ & $0.032$ \\
$150^{\circ}$ & $5.8$ & $3.376$ & $0.062$ \\
$160^{\circ}$ & $5.5$ & $5.839$ & $0.118$ \\
\end{tabular}
\caption{Maximum numbers of discernible signal photons $N_{\rm Sig}({\cal A}_{{\rm d},\mu})$ for different collision angles $\vartheta_{\rm col}$. The maximum is reached at $\mu=\mu_{\rm max}^{\rm Sig}$. For comparison, the corresponding number of background photons $N_{\rm Bg}({\cal A}_{{\rm d},\mu})$ is also given here. See Fig. \ref{fig:circMollSigMultiTheta} for the laser parameters employed here.} \label{tab:circNSigMax}
\end{table}

\begin{table}
\begin{tabular}{|*{4}{*{1}{|c}|}|}
$\vartheta_{\rm col}$ & $\mu=\mu_{\rm max}^{\rm Bg}$ & $N_{\rm Sig}({\cal A}_{{\rm d},\mu})$ &  $N_{\rm Bg}({\cal A}_{{\rm d},\mu})$ \\ \hline
$100^{\circ}$ & $4.5$ & $0.169$ & $0.005$ \\
$110^{\circ}$ & $4.5$ & $0.309$ & $0.009$ \\
$120^{\circ}$ & $4.4$ & $0.524$ & $0.016$ \\
$130^{\circ}$ & $4.2$ & $0.838$ & $0.022$ \\
$140^{\circ}$ & $4.0$ & $1.367$ & $0.041$ \\
$150^{\circ}$ & $4.0$ & $2.669$ & $0.082$ \\
$160^{\circ}$ & $3.8$ & $4.340$ & $0.161$ \\
\end{tabular}
\caption{Maximum numbers of background photons $N_{\rm Bg}({\cal A}_{{\rm d},\mu})$ for different collision angles $\vartheta_{\rm col}$. The maximum is reached at $\mu=\mu_{\rm max}^{\rm Bg}$. For comparison, the corresponding number of signal photons $N_{\rm Sig}({\cal A}_{{\rm d},\mu})$ is also given here. See Fig. \ref{fig:circMollSigMultiTheta} for the laser parameters employed here.}\label{tab:circNBgMax}
\end{table}

Tables \ref{tab:circNSigMax} and \ref{tab:circNBgMax} list the numbers of signal and background photons per shot at the corresponding values $\mu_{\rm max}^{\rm Sig}$ and $\mu_{\rm max}^{\rm Sig}$, respectively. 

A comparison of these values confirms that the number of discernible signal photons increases with $\vartheta_{\rm col}$.
Note, however, that the infinite Rayleigh range approximation allows for reliable insights only as long as 
$2/\left(\mu w_0\omega_0\right)<\left|\sin\vartheta_{\rm col}\right|$; cf. also Sect.~\ref{sec:exp}. 

Interestingly, the area of the regions $\mathcal{A}_{\rm d,\mu}$ decreases with increasing angle $\vartheta_{\rm col}$ -- except for the results with collision angle $\vartheta_{\rm col}=100^{\circ}$. 
For example, with $\vartheta_{\rm col}=130^{\circ}$ at $\mu_{\rm max}^{\rm Sig}=6.4$ we find $A(\mathcal{A}_{\rm d,6.4})\approx 1.10\pi$, whereas for $\vartheta_{\rm col}=160^{\circ}$ we have $A(\mathcal{A}_{\rm d,6.4})\approx 0.85\pi$. 
This indicates that the strength of the signal is more important than the size of the region where the signal surpasses the background.

\subsection{Probe with elliptical cross section}

In a next step we turn to a probe beam with an elliptical cross section. 
To this end, we introduce two independent waist sizes $w_x=\mu_xw_0$ and $w_y=\mu_yw_0$ for the probe beam. 
These values determine the two semi-axes of the elliptic cross section.

From the analysis of circular cross sections, see Section \ref{sec:circ}, we 
know that the signal becomes maximal for 
$w_x=w_y=\mu_{\rm max}^{\rm Sig} w_0$.
The tighter the focusing, the larger the intensity, which results in a stronger signal.
However, at the same time the divergence of the beam is increased, so that the discernible signal generically decreases.
If this beam is now focused harder along only the one of the semi-axes, the 
signal is still expected to increase, whereas it remains discernible in the 
perpendicular direction characterized by keeping the value $\mu_{\rm max}^{\rm 
Sig}$ fixed. 
Keeping other parameters fixed, such an elliptical cross section is expected to 
result in a larger yield of discernible signal photons $N_{\rm Sig}$ 
\cite{Karbstein:2015xra, Karbstein:2016lby}.

For generic choices of $\mu_x$ and $\mu_y$, the parameter space is significantly 
increased in comparison to the rotationally symmetric case 
characterized by a single parameter $\mu=\mu_x=\mu_y$.
For convenience and for simplifying the following discussion, we fix the 
parameter $\mu_x$ to the value maximizing the signal in section \ref{sec:circ} 
for the corresponding collision angle $\vartheta_{\rm col}$, i.e. set 
$\mu_x=\mu_{\rm max}^{\rm Sig} (\vartheta_{\rm col})$, cf. table 
\ref{tab:circNSigMax}. 
We choose $\mu_y(\vartheta_{\rm col})\in[1,\mu_{\rm max}^{\rm 
Sig}(\vartheta_{\rm col})]$ as a free parameter, since this leads to the 
maximum possible value of discernible signal photons at $\vartheta_{\rm 
col}=160^{\circ}$. 
For comparison, we discuss afterwards the case with $\mu_y=\mu_{\rm 
max}^{\rm Sig}$ and variable $\mu_x\in[1,\mu_{\rm max}^{\rm Sig}(\vartheta_{\rm 
col})]$.

In a first step, we focus on a collision of angle $\vartheta_{\rm col}=160^{\circ}$.
We set $\mu_x=\mu_{\rm max}^{\rm Sig}(160^{\circ})=5.5$, maximizing the 
discernible signal for the circular case, and keep $\mu_y$ as a free parameter. 
All other parameters remain unchanged. 

\begin{figure}[t]
	\centering
\includegraphics[width=\columnwidth]{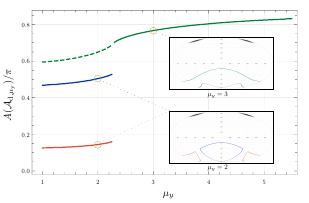}
	\caption{Area of solid angle regime $\mathcal{A}_{\rm d, \mu_y}$ as function of $\mu_y$ for a collision angle of $\vartheta_{\rm col}=160^{\circ}$. For low values of $\mu_y$ the region $\mathcal{A}_{\rm d \mu_y}$ is separated into two parts, as illustrated by the red and blue colored curves. Their sum yields the dashed green curve. The inlays show Mollweide plots of the areas for the example values of $\mu_y=2$ and $\mu_y=3$. For $\mu_y=3$ the region is simply connected; see the green solid line in the corresponding plot. Both pulses have the same frequency $\omega_0=1.55\,{\rm eV}$, energy $W=25\,{\rm J}$ and duration $\tau=25\,{\rm fs}$ (FWHM). Here, we depict results for a probe with an elliptical cross section ($w_x=5.5w_0$, $w_y=\mu_yw_0$) and a pump focused to $w_0=2\pi/\omega_0$.} 
	\label{fig:ellAd160} 
\end{figure}
First we consider the angular regions $\mathcal{A}_{\rm d, \mu_y}$ where the signal dominates the background.
Here the index $\mu_y$ indicates the parameter to be varied.

In Fig.~\ref{fig:ellAd160} we study the discernible solid angle area $A(\mathcal{A}_{\rm d, \mu_y})$ as function of $\mu_y$. 
Unlike before, there exist two distinct angular regions
for small values of $\mu_y$, where the signal becomes discernible.
This is illustrated in Fig.~\ref{fig:ellAd160} by the red and blue curves 
representing the areas of these regions; the red (blue) curve measures the area 
around $\varphi=0$ ($\varphi=180^{\circ}$). 
We denote the former (latter) region by $\mathcal{A}_{\rm d, \mu_y}^{(1)}$ ($\mathcal{A}_{\rm d, \mu_y}^{(2)}$).
In Fig.~\ref{fig:ellAd160} the third, dashed green curve indicates the sum of 
both areas, $A(\mathcal{A}_{\rm d, \mu_y}^{(1)}\cup\mathcal{A}_{\rm d, 
\mu_y}^{(2)}) = A(\mathcal{A}_{\rm d, \mu_y})$. For $\mu_y\approx2.3$ the two 
regions unite and beyond this value only a single angular region where the 
signal is discernible persists; cf. the green solid line.
Interestingly, the value of $\mu_y$ where the two regions merge amounts to an 
inflection point in the angular area. 
Figure \ref{fig:ellAd160} also shows two representative Mollweide projections of 
the boundaries of these regions for $\mu_y=2$ and $\mu_y=3$. 

Tracing these areas in Mollweide projections as a function of $\mu_y$, it can 
be observed how the two regions $\mathcal{A}_{\rm d, \mu_y}^{(1)}$ and 
$\mathcal{A}_{\rm d, \mu_y}^{(2)}$ slowly converge and finally merge.
In this process the upper bounds $\vartheta_{\rm u}^{(m)}(\varphi)$ do not 
change significantly; only the lower bounds $\vartheta_{\rm l}^{(m)}(\varphi)$ 
visibly change with $\mu_y$ predominantly in the vicinity of 
$\varphi=90^{\circ}$ and $\varphi=270^{\circ}$. 

Now, we compare the properties of the two disjoint regions for the example value of $\mu_y=2$. 
It is obvious that the area $A(\mathcal{A}_{\rm d,2}^{(1)})$ is much larger than 
$A(\mathcal{A}_{\rm d,2}^{(2)})$; $A(\mathcal{A}_{\rm 
d,2}^{(2)})/A(\mathcal{A}_{\rm d,2})\approx 22.3\%$. 
Interestingly, only $3.0\%$ of the background photons and $0.5\%$ of the signal 
photons are located in this $22.3\%$ of the total area. 
In absolute numbers, we count around 0.02 signal photons per shot in 
$\mathcal{A}_{\rm d,2}^{(2)}$ and 7.22 in $\mathcal{A}_{\rm d,2}^{(1)}$. 
With rotationally symmetric focusing, cf. section \ref{sec:circ} and in 
particular Fig.~\ref{fig:circMollSig}, the signal at $\varphi=0^{\circ}$ is 
significantly weaker compared to $\varphi=180^{\circ}$. 
An analogous behavior is observed for beams with elliptical average,  
the effect is even amplified by the different beam waists. 

\begin{figure*}[t]
	\centering
\includegraphics[width=\textwidth]{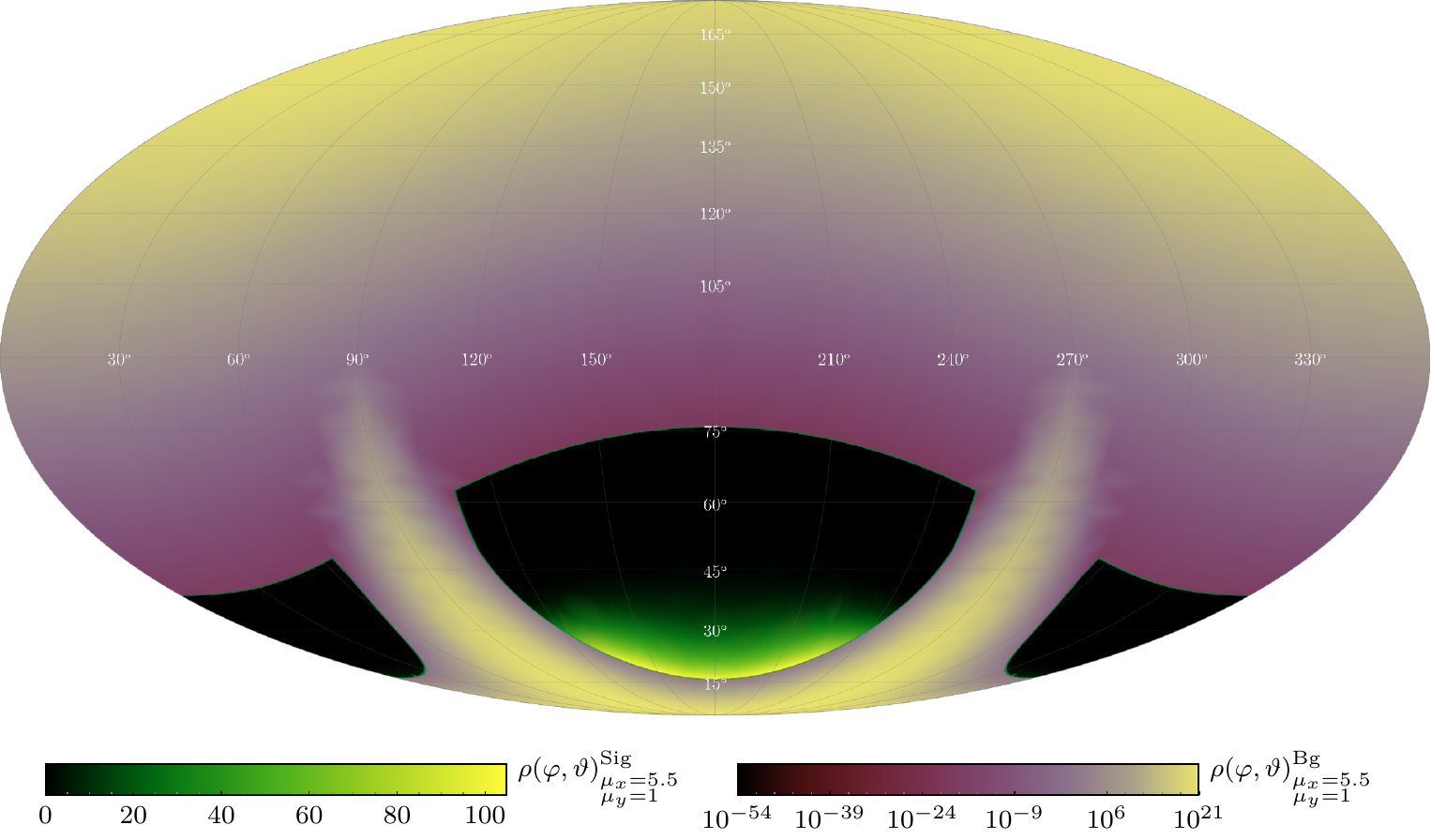}
	\caption{Mollweide plot (longitude $\varphi$, latitude $\vartheta$) of the signal $\rho_{1}^{\rm Sig}(\varphi,\vartheta)$ and background $\rho_{1}^{\rm Bg}(\varphi,\vartheta)$ photon densities for $\vartheta_{\rm col}=160^{\circ}$. The angular region where the signal is discernible is highlighted by solid green lines (left color scales). Outside these regions, the background dominates (right color scale). Both lasers deliver pulses of energy $W=25\,{\rm J}$ and duration $\tau=25\,{\rm fs}$ (FWHM) at a frequency of $\omega_0=1.55\,{\rm eV}$. The pump (probe) has a circular (elliptical) cross section. The pump waist is  $w_0=2\pi/\omega_0$ and the probe waists are $w_x=5.5w_0$ and $w_y=\mu_yw_0$, respectively.} 
	\label{fig:ellMollSigLog}
\end{figure*}
\begin{figure*}[]
	\centering
	\includegraphics[width=\textwidth]{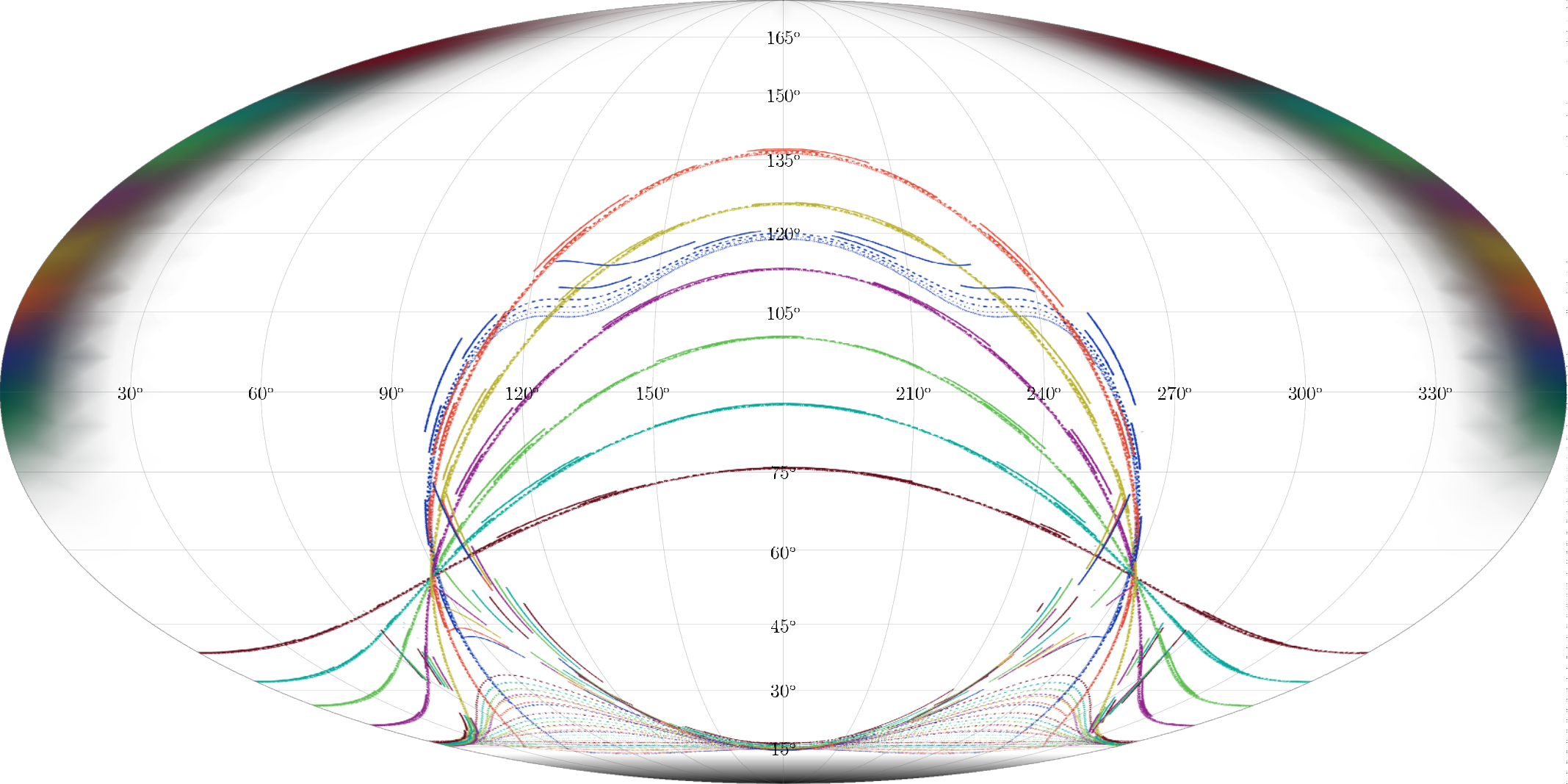}
	\caption{Mollweide plot (longitude $\varphi$, latitude $\vartheta$) showing angular regions where the signal is discernible for different collision angles $\vartheta_{\rm col}$: $100^{\circ}$ to $160^{\circ}$ in $10^{\circ}$ steps labeled by the color order: blue, red, yellow, magenta, light green, cyan, and brown. Additionally, the probe laser is marked by the black shade. This waist sizes of the elliptically focused probe  $w_x=\mu_{\rm max}^{\rm Sig}(\vartheta_{\rm col})\, 2\pi/\omega_0$ and $w_y=\frac{1}{5}(5+l(\mu_{\rm max}^{\rm Sig}(\vartheta_{\rm col})-1))\,\pi/\omega_0$. The contours for which $\rho_{\rm Sig}=\rho_{\rm Bg}$ holds for a given collision angle are marked by the differently colored lines associated with different values of $\mu_y=\frac{1}{5}(5+l(\mu_{\rm max}^{\rm Sig}(\vartheta_{\rm col})-1))$ with $l\in\mathbb{N}^{[0,5]}$: encoded by the type of lines; with higher wideness parameter, the spacing of the dashes is smaller. The driving laser pulses have the same energy $W=25\,\rm{J}$, duration $\tau=25\,{\rm fs}$, and frequency $\omega_0=1.55\,{\rm eV}$. The pump waist is $w_0=2\pi/\omega_0$.}
	\label{fig:ellMollSigMultiTheta}
\end{figure*}

Figure \ref{fig:ellMollSigLog} illustates the densities $\rho_{1}^{\rm 
Sig}(\varphi,\vartheta)$ and $\rho_{1}^{\rm Bg}(\varphi,\vartheta)$ for 
the example  $\mu_x=5.5$, $\mu_y=1$ and $\vartheta_{\rm 
col}=160^{\circ}$. 
In contrast to Fig.~\ref{fig:circMollSig}, we use here a logarithmic scale for 
the background.
Figure \ref{fig:ellMollSigLog} clearly shows how the elliptical probe cross section affects the discernible signal.
The ocher colored area in the lower half sphere reflects the elliptical probe cross section.
Its shape is reminiscent to a strongly curved banana, which overlays parts of 
the region of the signal. 
In the region around $\vartheta=180^{\circ}$ the signal is much stronger than in the region around $\varphi=0$ separated from it.
This is consistent with previous observations at $\mu_y=2$: the former region 
accounts for $21.2\%$ of the total area and contains only $3\%$ of the 
background and $0.5\%$ of the discernible signal photons. 
Considering the union of both regions, where the signal dominates the 
background, 11.63 signal photons per shot can be counted, and 0.25 background 
photons per shot.

Figure \ref{fig:ellMollSigMultiTheta} shows the regions of the dominant signals 
for different collision angles $\vartheta_{\rm col}$.
The parameter $\mu_x=\mu_{\rm max}^{\rm Sig}(\vartheta_{\rm col})$ is fixed for 
each collision angle $\vartheta_{\rm col}$ to a corresponding value listed in 
table \ref{tab:circNSigMax}. 
At the same time, the parameter $\mu_y$ is varied in the interval $[1,\mu_{\rm max}^{\rm Sig}(\vartheta_{\rm col})]$.
Figure \ref{fig:ellMollSigMultiTheta} clearly shows that the upper bound 
$\vartheta_{\rm u}(\varphi)$ typically does not change appreciably with $\mu_y$ 
for constant $\vartheta_{\rm col}$.
The only exception is the angle $\vartheta_{\rm col}=100^{\circ}$. 
Here the collision angle is close to $\vartheta_{\rm col}\approx90^{\circ}$; 
thus the influence of the width of the probe beam on the upper bond 
$\vartheta_{\rm u}(\varphi)$ varies more, cf. also 
Fig.~\ref{fig:circMollSigMultiTheta}. 

At the same time the lower boundary $\vartheta_{\rm l}(\varphi)$ changes strongly with $\vartheta_{\rm col}$.
Here we observe that pronounced local maxima around $\varphi=90^{\circ}$ and 
$\varphi=270^{\circ}$ appear for decreasing $\mu_y$.
If the value of $\vartheta_{\rm l}(\varphi)$ at the maxima reaches the 
corresponding upper bound $\vartheta_{\rm u}(\varphi)$, then two separate 
angular regions are formed. 
However, this effect appears only at larger collision angles -- starting from 
$\vartheta_{\rm col}\simeq120^{\circ}$ -- and becomes more pronounced with 
larger angles $\vartheta_{\rm col}$.  

The discernible-signal area increases with $\mu_y$ for each $\vartheta_{\rm 
col}$.
At larger collision angles $\vartheta_{\rm col}$ the total area 
$A(\mathcal{A}_{\rm d,\mu_y})$ is smaller than for smaller $\vartheta_{\rm 
col}$. 

\begin{figure}[h!]
	\centering
	\includegraphics[width=\columnwidth]{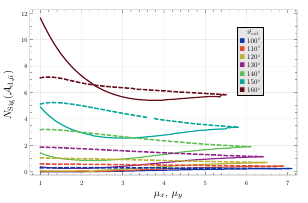}
	\caption{Discernible number of signal photons in the solid angle regime $\mathcal{A}_{\rm d, \bar{\mu}}$ as function of $\mu_y$ ($\mu_x$) with constant $\mu_x=\mu_{\rm max}^{\rm Sig}$ ($\mu_y=\mu_{\rm max}^{\rm Sig}$) for different collision angles $\vartheta_{\rm col}$ illustated by solid (dashed) lines. Both lasers deliver frequency $\omega_0=1.55\,{\rm eV}$ photons at a pulse energy  $W=25\,{\rm J}$ and duration $\tau=25\,{\rm fs}$ (FWHM). The pump is focused to  $w_0=2\pi/\omega_0$. The probe waists are $w_x=\mu_{\rm max}^{\rm Sig}\,w_0$ and $w_y=\mu_yw_0$ ($w_x=\mu_xw_0$ and $w_y=\mu_{\rm max}^{\rm Sig}\,w_0$).}
	\label{fig:ellSigMultiTheta}
\end{figure}

In Fig.~\ref{fig:ellSigMultiTheta}, we highlight the dependence of 
the number of discernible signal photons $N_{\rm Sig}(\mathcal{A}_{\rm d, 
\mu_y})$on $\mu_y$ for different collision angles by solid lines.
Additionally, we mark the number of discernible signal photons 
as function of $\mu_x$ by dashed lines with a constant choice of 
$\mu_y=\mu_{\rm max}^{\rm Sig}$. 
Obviously, the discernible signal $N_{\rm Sig}(\mathcal{A}_{\rm d, \bar{\mu}})$ 
increases with increasing collision angle $\vartheta_{\rm col}$ for each value 
of $\mu_y$ and $\mu_x$, respectively.

First we consider the results with fixed $\mu_x$ and varying $\mu_y$.
It is interesting to note that two (local) maxima can occur in the $\mu_y$ intervals studied. 
In any case the value $N_{\rm Sig}(\mathcal{A}_{\rm d, \mu_{\rm max}^{\rm 
Sig}})$ for $\mu_y=\mu_{\rm max}^{\rm Sig}$ is a maximum; cf. also discussion in 
section \ref{sec:circ}. 
For all graphs with $\vartheta_{\rm col}>100^{\circ}$ and variable $\mu_y$, 
there exists a local minimum in the interval $\mu_y\in[1,\mu_{\rm max}^{\rm 
Sig}]$. 
Therefore, the function increases with $\mu_y$ smaller than $\mu_y$ at this local minimum.
This leads to a maximum value (not necessarily a local maximum) at the 
position $\mu_y=1$.
However, this value is only for the collision angles $\vartheta_{\rm 
col}=150^{\circ}$ and $\vartheta_{\rm col}=160^{\circ}$ larger than the signal 
photon number at $\mu_y=\mu_{\rm max}^{\rm Sig}$. 
This implies that for smaller collision angles $\vartheta_{\rm col}<150^{\circ}$ 
elliptical cross sections do not result in an increased discernible 
signal as long as the semi-axes $\mu_y<\mu_{\rm max}^{\rm Sig}$ and 
$\mu_x=\mu_{\rm max}^{\rm Sig}$ are fixed. 
On the other hand, for large collision angles $\vartheta_{\rm col}$ elliptical probe cross sections can increase the discernible signal. 
For $\vartheta_{\rm col}=150^{\circ}$ we determine $N_{\rm Sig}(\mathcal{A}_{\rm 
d,1})=4.90$ signal photons per shot, for $\vartheta_{\rm col}=160^{\circ}$ it is 
even $N_{\rm Sig}(\mathcal{A}_{\rm d,1})=11.63$. 

If, on the other hand, we consider $\mu_x$ as a free parameter and fix 
$\mu_y=\mu_{\rm max}^{\rm Sig}$ for the corresponding angle $\vartheta_{\rm 
col}$, we find in all cases that the number of discernible signal photons 
increases with decreasing $\mu_x$.  
Except for $\vartheta_{\rm col}=140^{\circ}$, $\vartheta_{\rm col}=150^{\circ}$ 
and $\vartheta_{\rm col}=160^{\circ}$, the maximum value is always found at 
$\mu_x=1$ and $\mu_y=\mu_{\rm max}^{\rm Sig}$, see Tab.~\ref{tab:ellNSigMax}.  
Moreover, except for the angle $\vartheta_{\rm col}=160^{\circ}$, the maxima are 
always larger than in the case of variable $\mu_y$ and fixed $\mu_x$. With 
decreasing collision angle $\vartheta_{\rm col}$, this is because  the 
pronounced maxima of $\rho(\varphi,\vartheta_{\rm l}(\varphi))$ of the signal 
are closer to $\varphi=0$ or $\varphi=180^{\circ}$, i.e. the collision plane. 
Only at the angle $\vartheta_{\rm col}=160^{\circ}$ with $\varphi=114.5^{\circ}$ 
and $\varphi=245.5^{\circ}$ the pronounced maxima are closer to the plane of 
$w_y$ than $w_x$. 
At the same time, the maximum of $N(\mathcal{A}_{{\rm d},\bar{\mu}})$ for 
$\vartheta_{\rm col}=160^{\circ}$ is significantly larger for $\mu_y=1$, 
$\mu_x=\mu_{\rm max}^{\rm Sig}$, which is why we focus here on the case of 
variable $\mu_y$. 

The kinks visible in some graphs in Fig.~\ref{fig:ellSigMultiTheta}
are numerical artifacts; they become more pronounced as the collision 
angle $\vartheta_{\rm col}$ increases.
They occur when two regions merge, as shown in Fig.~\ref{fig:ellAd160}, 
affecting the precision of the numerical integration.
Also, the artifacts appear at the value for $\mu_y$ where $\phi_i=0$ and 
$\varphi_f=2\pi$ for the boundary functions $\vartheta_{\rm l/u}(\varphi)$ of 
$\mathcal{A}_{\rm d,\mu_y}$ appears for the first time. 

\begin{table}
\begin{tabular}{|*{5}{*{1}{|c}|}|}
$\vartheta_{\rm col}$ & $\mu_x$ & $\mu_y$ & $N_{\rm Sig}({\cal A}_{{\rm d},\bar{\mu}})$ &  $N_{\rm Bg}({\cal A}_{{\rm d},\bar{\mu}})$ \\ \hline
$100^{\circ}$ & $1$ & $7.1$ & $0.294$ & $0.009$ \\
$110^{\circ}$ & $1$ & $6.9$ & $0.586$ & $0.009$ \\
$120^{\circ}$ & $1$ & $6.5$ & $1.044$ & $0.016$ \\
$130^{\circ}$ & $1$ & $6.4$ & $1.864$ & $0.032$ \\
$140^{\circ}$ & $1.113$ & $6.1$ & $3.205$ & $0.061$ \\
$150^{\circ}$ & $1.32$ & $5.8$ & $5.243$ & $0.116$ \\
$160^{\circ}$ & $5.5$ & $1.$ & $11.633$ & $0.248$ \\
\end{tabular}
\caption{Maximum numbers of discernible signal photons $N({\cal A}_{{\rm d},\bar{\mu}})$ and its corresponding relative probe waist sizes  $\mu_y$ and $\mu_y$ for two-beam collision with elliptic probe for different collision angles $\vartheta_{\rm col}$. These parameters are selected according to the maximum signal, whereby at least one parameter fulfills $\mu_x=\mu_{\rm max}^{\rm Sig}(\vartheta_{\rm col})$ or $\mu_y=\mu_{\rm max}^{\rm Sig}(\vartheta_{\rm col})$. For comparison, the corresponding number of background photons $N_{\rm Bg}({\cal A}_{{\rm d},\bar{\mu}})$ is also given here. See \Figref{fig:ellSigMultiTheta} for the laser parameters employed here.}
\label{tab:ellNSigMax}
\end{table}

In table \ref{tab:ellNSigMax} we list the maximal discernible signal 
photon number for the parameters considered here. 
Only for $\vartheta_{\rm col}=160^{\circ}$, the maxmimum is detected for the 
fixed choice $\mu_x=\mu_{\rm max}^{\rm Sig}(\vartheta_{\rm col})$. 
Since this signal is much larger than the other maximum, we limit our 
discussion to this parameter choice.
In addition, table \ref{tab:ellNSigMax} lists the corresponding number of background photons.
Here, we want to emphasize again that in the determined regions the signal 
significantly dominates the background. 
d thus this focus can contribute to a larger maximum background.

\subsection{Quantum vacuum birefringence} \label{sec:qBi}

Let us finally consider the phenomenon of birefringence.
For this, we concentrate exclusively on the collision angle $\vartheta_{\rm 
col}=160^{\circ}$ and the two parameter sets of circular or
elliptical cross sections where polarization insensitive measurements maximize 
the discernible signal, i.e. we use the width parameters $\mu=5.5$ or 
$\mu_x=5.5$, $\mu_y=1$, respectively.

For birefringence, we now set the polarization angle $\beta_2$ of the pump beam 
to $\beta_2=45^{\circ}$, as explained earlier in the section \ref{sec:exp}.
We consider only signal photons whose polarization is perpendicular to the 
original linear polarization of the probe beam with flipped polarization angle 
$\beta_{\perp}$ \cite{Dinu:2013gaa,Karbstein:2015xra}.
This implies ${\bf e}_{\beta_{\perp}}\cdot {\bf E}_1({\bf x},t)=0$ such that 
$\beta_{\perp}=\arctan\left(\cos\vartheta\cot\varphi\right)$ holds.

We determine the number of signal photons as before with the restriction that 
no sum over the polarizations is performed. Instead, we use $(p)\rightarrow 
\beta_{\perp}$ and derive the corresponding signal photon density
\begin{align}
 \rho_{\bar{\mu}}^{\rm Sig,\perp}(\varphi,\vartheta) &=\frac{1}{(2\pi)^7} \frac{2}{45^2} \left(\frac{e}{m_e}\right)^8 \nonumber \\
 &\phantom{=}\times \Big[  g_{\beta_{\perp}\frac{\pi}{4},121}^2(\varphi,\vartheta) \!\int_{-\infty}^{\infty} \!\!\!\mathrm{dk}\,{\rm k}^3\, \mathcal{I}_{121}^2({\bf k})   \nonumber \\
  &\phantom{=\times} +  g_{\beta_{\perp}\frac{\pi}{4},212}^2(\varphi,\vartheta) \!\int_{-\infty}^{\infty} \!\!\!\mathrm{dk}\,{\rm k}^3\, \mathcal{I}_{212}^2({\bf k}) \Big] \,. \label{eq:rhoSigmubarperp}
\end{align}
Compared with Eq.~\ref{eq:rhoSigmubar}, only the value of the function 
$g_{\beta_{\perp}\frac{\pi}{4},iji}$ changes.
The number of polarization flipped signal photons $N_{\rm 
Sig,\perp}(\mathcal{A},\bar{\mu})$ follows analogously to the Eq.~\ref{eq:Nsig}.

\begin{figure}[t]
	\centering
	\includegraphics[width=\columnwidth]{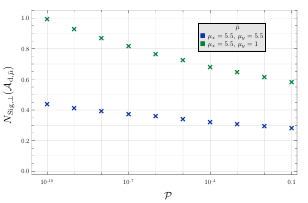}
	\caption{Discernible number of polarization flipped signal photons in the solid angle regime $\mathcal{A}_{\rm d, \bar{\mu}}$ as function pof the purity $\mathcal{P}$ of the polarization of the probe beam in logarithmic scale. Here only the collision scenarios for a collision angles $\vartheta_{\rm col}=160^{\circ}$ are analyzed. Both linear polarized driving lasers deliver frequency $\omega_0=1.55\,{\rm eV}$ photons at a pulse energy  $W=25\,{\rm J}$ and duration $\tau=25\,{\rm fs}$ (FWHM). The pump is focused to  $w_0=2\pi/\omega_0$. The probe waists are $w_x=\mu_x\,w_0$ and $w_y=\mu_yw_0$.}
	\label{fig:Birefringence}
\end{figure}

In order to estimate the background, we introduce the polarization purity 
$\mathcal{P}$ characterizing the quality of the probe polarimetry 
polarizer-analyzer system \cite{Bernhardt:2020vxa}.
Here we consider realistic purities on the range from 
$\mathcal{P}=10^{-1}$ to $\mathcal{P}=10^{-10}$ in logarithmic steps.
For a conservative estimate of the solid angle region of the discernible 
signal, we use
\begin{align}
 \mathcal{A}_{\rm d,\bar{\mu}} &= \Big\{ (\varphi,\vartheta)\in\left[0,2\pi\right]\times\left[0,\pi\right) \left.\right|\nonumber \\
 & \phantom{= \Big\{} \rho^{\rm Sig,\perp}_{\bar{\mu}}(\varphi,\vartheta) \geq \mathcal{P}\frac{{\rm d}N^{\rm Bg}_1}{{\rm d}\Omega} + \frac{{\rm d}N^{\rm Bg}_2}{{\rm d}\Omega}  \Big\}\,,
\end{align}
where we have not applied any polarization constraints to the background of the 
pump beam. The latter has no influence in the relevant region of the flipped 
signal and additionally gives us an upper bound $\vartheta_{\rm u}(\varphi)$ for 
easier evaluation of the integrations.

In Fig.~\ref{fig:Birefringence}, the number of signal photons in the dominant 
regions $N_{\rm Sig,\perp}( \mathcal{A}_{\rm d,\bar{\mu}})$ is plotted as 
a function of the purity $\mathcal{P}$. 
We also distinguish between an elliptical and a circular probe beam. 
In both cases there is always a sufficient number of discernible signal photons; 
as expected the number decreases with the purity, however, only rather 
mildy in accordance with the findings of \cite{Karbstein:2021ldz}. 

Let us discuss the results for $\mathcal{P}=10^{-10}$.
Considering a probe beam with circular cross section we obtain $N_{\rm 
Sig,\perp}(\mathcal{A}_{\rm d,5.5})= 0.44$ signal photons per shot against a 
background of $N_{\rm Bg,\perp}(\mathcal{A}_{\rm d,5.5})= 0.01$ on an area of 
$A(\mathcal{A}_{\rm d,5.5})= 0.85\pi$. 
In the case of the elliptically focused probe beam we get $N_{\rm 
Sig,\perp}(\mathcal{A}_{\rm d,1})= 1.00$ discernible signal photons per shot and 
a background of $N_{\rm Bg,\perp}(\mathcal{A}_{\rm d,1})= 0.03$. 
The area of the region of the discernible signal is $A(\mathcal{A}_{\rm d,1})= 0.66\pi$. 
In order to put these values into context, let us 
compare them with the number of signal photons in a forward cone  
$\mathcal{A}_{\frac{\pi}{4}}=\left\{(\varphi,\vartheta)\in\left[0,2\pi\right]
\times\left[0,\frac{\pi}{4}\right]\right\}$.  
Here we count $N_{\rm Sig,\perp}(\mathcal{A}_{\frac{\pi}{4}})=0.86$ by using a 
probe beam with circular cross section or $N_{\rm 
Sig,\perp}(\mathcal{A}_{\frac{\pi}{4}})=3.67$ with an elliptical probe. 
Accordingly, it is possible to measure a significant fraction $0.51\%$ 
($0.27\%$) of the whole signal for a circular (elliptical) probe. 

However, the number of driving laser photons beyond the output polarity 
occurring in $\mathcal{A}_{\frac{\pi}{4}}$ exceeds these numbers by far ($N_{\rm 
Bg,\perp}(\mathcal{A}_{\frac{\pi}{4}})\approx 10^{20}$), which shows that our 
method for optimizing the measurement region of the signature of the quantum 
vacuum is also useful for the effect of birefringence. 

The rather weak dependence of the discernible-photon number on the 
polarization purity might seem surprising but is completely consistent with the observations of vacuum birefringence studies with x-ray probes such as, e.g., \cite{Schlenvoigt:2016,Karbstein:2021ldz,Mosman:2021vua}: while a substantial degree of 
polarization purity is absolutely essential for a birefringence measurement on 
axis, a less pure setup can be compensated for by a scenario that includes 
scattering into low-background regions.

\section{Conclusions and Outlook}\label{sec:conc}

In an effort to increase the signal of nonlinear vacuum-induced 
interactions, we have studied the relevance of the choice of beam waists in the 
collision of two optical laser pulses. Focusing on collision angles in the 
range $100^{\circ}\leq\vartheta_{\rm col}\leq 160^{\circ}$, we see a decisive 
dependence of the detectable signal strength in terms of \textit{discernible} 
photons on the beam waist of the probe beam both for circular as well as 
elliptical beam cross section.

In order to simplify the experimental scenario, we consider the two 
pulses as originating from the same ultra-intense laser source, having similar 
properties with respect to pulse duration, frequency, and total energy; the 
common source is taken to resemble the parameters available in 
\cite{Doyle:2021mdt,Hartmann:2021iyf,CALA,JETI-200}.
For a mostly analytically accessible and efficient theoretical modeling, the 
laser pulses are considered in the infinite Rayleigh range approximation.
This allows for a controlled determination and analysis of discernible signals 
and their angular emission regimes in the considered parameter ranges.

The essential mechanisms become already visible for the simpler case of a 
probe beam with circular cross section. One of our key findings is that 
the maximization of the discernible signal requires a comprise between 
increasing the intensity in the collision region and decreasing the background 
in the detection region. While the signal increases quadratically with the 
intensity, the background decreases exponentially beyond the outgoing laser 
cone which in turn decreases for larger waist sizes. In the range of collision 
angles $100^{\circ}<\theta_{\rm col}<160^{\circ}$, we observe that an optimal 
beam waist is about $\sim 5 \dots 7$ times bigger than the diffraction limit. 
Our results are in line with those found in a counterpropagating setup 
\cite{Mosman:2021vua}.

For the largest scattering angle studied in this work, elliptical cross 
sections come with further mechanisms to increase the signal: suitable choices 
of the ellipticity can increase the intensity in the collision region while 
maintaining the background suppression in the detection region.

Additionally, we investigated the discernible photon signal including a 
polarization flip in promising laser configurations. 
This signal of birefringence was determined as a function of the 
polarization purity; the results are promising for probe beams with circular 
cross section as well as with elliptical cross section.

For completeness, let us point out that further sources of background 
photons can become relevant in experiments. 
Apart from the photons comprising the driving laser beams included in 
our present study, there is generically a scattering background.
Such scattering originates in diffration of laser photons off the walls 
of the vacuum chamber and optical apertures both upstream from the beam 
guidance as well as downstream in the collection optics.
Furthermore  (relativistic) Thomson and (relativistic) Rayleigh scattering by 
residual gas ions and their electrons can occur and need to be controlled and 
reduced in experiment. First ideas of reducing these 
background sources through space and time resolved 
detection have already been successfully implemented \cite{Doyle:2021mdt}.

We expect that the mechanisms 
studied in this work that maximize the signal can also be operative in 
collisions of a larger number of pulses \cite{Klar:2020ych, Moulin:1999hwj, Lundstrom:2005za, Lundin:2006wu, Gies:2017ezf, Aboushelbaya:2019ncg}; in addition, 
also scattering into different frequency channels can lead to further 
enhancements in such scenarios \cite{Fillion-Gourdeau:2014uua, Huang:2019ojh, Gies:2021ymf, Sasorov:2021anc, Sangal:2021qeg}.

In summary, the present work demonstrates that a 
detailed modeling of laser pulse collisions based on theory from first 
principles can identify unexpected sweet spots for future discovery experiments 
of QED vacuum nonlinearities.

\acknowledgments

We are grateful to Richard Schmieden for helpful discussions.
This work has been funded by the Deutsche Forschungsgemeinschaft (DFG) under 
Grant Nos. 392856280, 416607684, and 416611371 within the Research Unit FOR2783/2.

\begin{appendix}

\section{Solid angle regions of discernible signal} \label{sec:Ad}

\begin{figure*}[t]
	\centering
	\includegraphics[width=\textwidth]{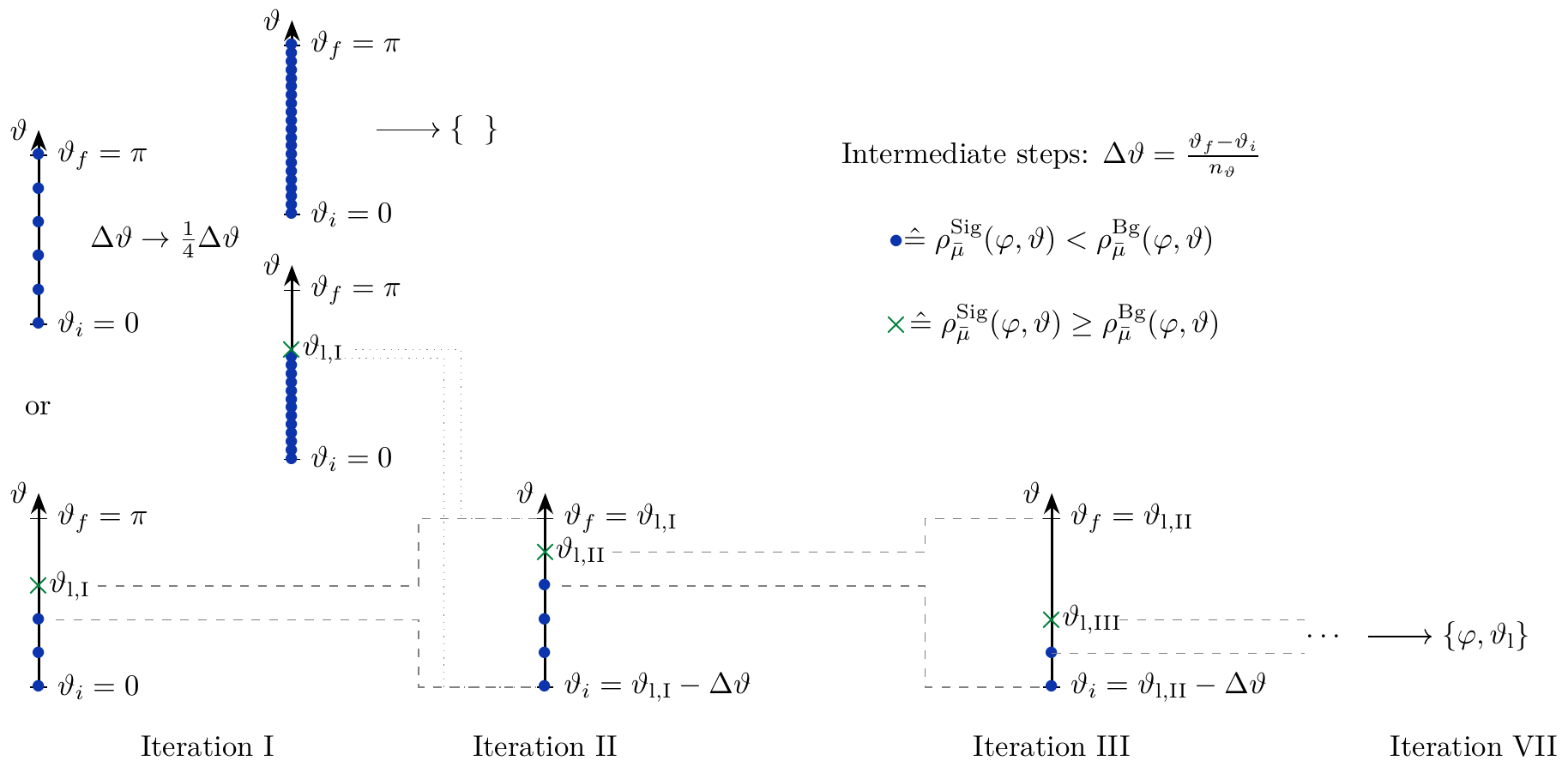}
	\caption{Scheme for the numerical determination of the lower bound of 
$\vartheta_{\rm l}$ at constant $\varphi$ for a dominant signal. The evaluated 
functions $\rho_{\bar{\mu}}^{\rm Sig}(\varphi,\vartheta)$ and 
$\rho_{\bar{\mu}}^{\rm Bg}(\varphi,\vartheta)$ are compared at fixed $\varphi$ 
on the interval $[\vartheta_i,\vartheta_f]$, starting in steps $\Delta\vartheta$ 
ascending from $\vartheta_i$. In the first iteration $s={\rm I}$ we define 
$\vartheta_i\equiv0$ and $\vartheta_f\equiv\pi$. If always 
$\rho_{\bar{\mu}}^{\rm Bg}(\varphi,\vartheta)>\rho_{\bar{\mu}}^{\rm 
Sig}(\varphi,\vartheta)$ is valid, the interval is scanned again with a quarter 
of the step size. If there is no finding for given $\varphi$ the signal is 
classified as recessive in this region. If the signal is dominant at 
$\vartheta_{{\rm l},s}$, then the run is terminated prematurely and the next 
iteration step $s\rightarrow s+1$ begins. Here, the interval boundaries are 
adjusted to $\vartheta_i\equiv\vartheta_{{\rm l},s-1}-\Delta\vartheta$ and 
$\vartheta_f\equiv\vartheta_{{\rm l},s-1}$ according to the previous iteration 
step. The step size is redetermined and the procedure starts again. After seven 
iterations, the result $\vartheta_{\rm l}=\vartheta_{\rm l,VII}$ is used.} 
	\label{fig:Iteration_Theta_d}
\end{figure*}

A signal is discernible if the local photon density of the 
signal surpasses the density of the photons of the driving lasers. 
The solid angle region where the signal dominates is defined by
\begin{equation}
 \mathcal{A}_{\rm d,\bar{\mu}} = \left\{ 
(\varphi,\vartheta)\in\left[0,2\pi\right]\times\left[0,\pi\right) \left.\right| 
\rho^{\rm Sig}_{\bar{\mu}}(\varphi,\vartheta) \geq \rho^{\rm 
Bg}_{\bar{\mu}}(\varphi,\vartheta) \right\}\,. 
\end{equation}
The shape of this region depends on the parameters of the interacting lasers, in 
particular the choice of the beam waists, represented by $\bar{\mu}$, and the 
collision angle $\vartheta_{\rm col}$. 
This region can be of variable shape and is in general  not simply connected. 
If the signal is nowhere discernible, then $\mathcal{A}_{\rm 
d,\bar{\mu}}=\{\;\}$.

We denote the boundary of the region $\mathcal{A}_{\rm d,\bar{\mu}}$ by
\begin{equation}
 \partial\mathcal{A}_{\rm d,\bar{\mu}} = \left\{ (\varphi,\vartheta)\in\left[0,2\pi\right]\times\left[0,\pi\right) \left.\right| \rho^{\rm Sig}_{\bar{\mu}}(\varphi,\vartheta) = \rho^{\rm Bg}_{\bar{\mu}}(\varphi,\vartheta) \right\}\,.
\end{equation}

For simply connected regions, the boundary can be parameterized by two 
functions $\vartheta_{\rm l}(\varphi)\leq\vartheta_{\rm u}(\varphi)$ depending 
on the azimuthal angle $\varphi$, where $\vartheta_{\rm l}(\varphi)$ denotes the 
lower bound and $\vartheta_{\rm u}(\varphi)$ the upper bound for a discernible 
signal region. 
These functions are defined in the interval $[\varphi_i,\varphi_f]$ with 
$\varphi_i$ and $\varphi_f$ implicitly determined by $\vartheta_{\rm 
l}(\varphi_i)=\vartheta_{\rm u}(\varphi_i)$ and $\vartheta_{\rm 
l}(\varphi_f)=\vartheta_{\rm u}(\varphi_f)$ for $\varphi_i>0$ and 
$\varphi_f<2\pi$. 
If the latter condition is violated, $\vartheta_{\rm l/u}(0)=\vartheta_{\rm 
l/u}(2\pi)$ applies in both cases, where the values of the two functions on the 
upper and lower boundary may now be unequal. In this case, we have 
$\partial\mathcal{A}_{\rm d,\bar{\mu}} = \left\{\vartheta_{\rm 
l}(\varphi)\left.\right| \varphi\in\left[\varphi_i,\varphi_f\right] \right\}
\cup\left\{\vartheta_{\rm u}(\varphi)\left.\right| 
\varphi\in\left[\varphi_i,\varphi_f\right] \right\}$. 

If the region $\mathcal{A}_{\rm d,\bar{\mu}}$ is not simply connected, but at 
most two mappings to $\vartheta$ can be found for any given value of $\varphi$, 
then the $\varphi$ domain can be divided into $M$ connected intervals 
$[\varphi_i^{(m)}, \varphi_f^{(m)}]$ labeled by $m=1,2\dots,M$, 
where the lower and upper bounds $\vartheta_{l/u}^{(m)}(\varphi)$ can be 
defined such that $\partial\mathcal{A}_{\rm d,\bar{\mu}} = 
\bigcup_{m=1}^M\Big(\left\{\vartheta_{\rm l}^{(m)}(\varphi)\left.\right| 
\varphi\in\left[\varphi_i^{(m)},\varphi_f^{(m)}\right] \right\} 
\cup\left\{\vartheta_{\rm u}^{(m)}(\varphi)\left.\right| \varphi\in\left[\varphi_i^{(m)},\varphi_f^{(m)}\right] \right\}\Big)$. 
It is possible to discuss even more complicated shapes of $\mathcal{A}_{\rm 
d,\bar{\mu}}$ and $\partial\mathcal{A}_{\rm d,\bar{\mu}}$, respectively, but 
this is not necessary in the present case, cf. below. 

In order to determine $\mathcal{A}_{\rm d,\bar{\mu}}$ explicitly, we first 
calculate $\partial\mathcal{A}_{\rm d,\bar{\mu}}$ by determining all possible 
functions $\vartheta(\varphi)$, which solve the equation $\rho^{\rm 
Sig}_{\bar{\mu}}(\varphi,\vartheta) = \rho^{\rm 
Bg}_{\bar{\mu}}(\varphi,\vartheta)$ on 
$(\varphi,\vartheta)\in\left[0,2\pi\right]\times\left[0,\pi\right)$. 
Since $\rho^{\rm Sig}_{\bar{\mu}}$ in general exhibits a complicated dependency 
on $\vartheta$ and $\varphi$, we construct a algorithm for its 
determination.  

Being aware of potential shortcomings of this approach discussed below, we 
determine both a lower bound $\vartheta_{\rm l}(\varphi)$ and upper bound 
$\vartheta_{\rm u}(\varphi) \geq \vartheta_{\rm l}(\varphi)$ in a 
first step. 
For this end, we sample the interval $\varphi\in[0,2\pi]$ by $n_{\varphi}$ 
sampling points, usually using $n_{\varphi}=300$ or bigger. 
For each given value of $\varphi$, we then determine the lower/upper bounds 
$\vartheta_{\rm l/d}(\varphi)$ using the iterative procedure illustrated 
schematically in Fig.~\ref{fig:Iteration_Theta_d}.  
In iteration I the interval $\vartheta\in[0,\pi)$ is divided into 
$n_{\vartheta}$ points with $\Delta\vartheta=\frac{\pi}{n_{\vartheta}}$, using 
$n_{\vartheta}=70$ or bigger. 
Depending on whether we are looking for the upper or lower bound we start with 
$\vartheta=\pi$ or $\vartheta=0$. 
In the following we concentrate on the determination of the lower bound.
To this end, we evaluate both $\rho_{\bar{\mu}}^{\rm Sig}$ and 
$\rho_{\bar{\mu}}^{\rm Bg}$ at each sampling point with ascending value of 
$\vartheta\rightarrow\vartheta+\Delta\vartheta$ from the initial value 
$\vartheta=0$ and compare them.  
If $\rho^{\rm Sig}_{\bar{\mu}}(\varphi,\vartheta)<\rho^{\rm 
Bg}_{\bar{\mu}}(\varphi,\vartheta)$ holds, we increase $\vartheta$ by 
$\Delta\vartheta$ and compare again. 
If the signal density becomes larger or equal to the background density, we 
terminate the procedure and register this value as $\vartheta_{\rm l,I}$. 
If the first iteration did not result in a positive outcome, we decrease the 
step size $\Delta\vartheta$ by one-fourth, 
$\Delta\vartheta\rightarrow\frac{1}{4}\Delta\vartheta$, and repeat iteration I. 
If this also does not yield a result, we assume an empty solution for this value of $\varphi$. 
We continue with the next iteration step.  
 
In iteration II, we refine the outcome of iteration I by adjusting the bounds of 
the $\vartheta$ domain to be studied. 
As initial value, we now use $\vartheta_i=\vartheta_{\rm l,I}-\Delta\vartheta$, 
as final value $\vartheta_f=\vartheta_{\rm l,I}$. 
From these two values a new step size $\Delta\vartheta=\frac{\vartheta_f-\vartheta_i}{n_{\vartheta}}$ is calculated.
Starting with $\vartheta=\vartheta_i$ the values of $\rho_{\bar{\mu}}^{\rm Sig}$ 
and $\rho_{\bar{\mu}}^{\rm Bg}$ are compared again until the signal dominates. 
The corresponding value is registered and $\vartheta_{\rm l,II}$ is used as 
input for the next iteration III which resembles iteration II with 
$\vartheta_{\rm l,I}\rightarrow\vartheta_{\rm l,II}$. 

After seven iterations we stop and use $\vartheta_{\rm l, VII}=\vartheta_{\rm l}$.
By successively decreasing the intervals and step sizes, this value has a 
maximum error of $\pi n_{\vartheta}^{-7}$, which is negligible in the context of 
the accuracy of the applied approximations. 

Applying this procedure to all sampling points of $\varphi$ we obtain a table of 
mappings $\{\varphi,\vartheta_{\rm l}\}$. 
We use this table to interpolate a function $\vartheta_{\rm l}(\varphi)$ (or 
several functions for regions which are not simply connected). 
For this interpolation we fit third-order polynomials between successive 
data points. 
Analogously, we obtain the interpolation $\vartheta_{\rm u}(\varphi)$ of the upper bound.
These results allow us to determine the solid angle region where the signal 
dominates the background as $\mathcal{A}_{\rm d,\bar{\mu}}=\mathcal{A}_{\rm 
l,{\bar{\mu}}}\cap\mathcal{A}_{\rm u,{\bar{\mu}}}$ with $\mathcal{A}_{\rm 
l,{\bar{\mu}}}=\left\{ 
(\varphi,\vartheta)\in\left[0,2\pi\right]\times\left[0,\pi\right) \left.\right| 
\vartheta(\varphi)\geq \vartheta_{\rm l}(\varphi)) \right\}$ and 
$\mathcal{A}_{\rm u,{\bar{\mu}}}=\left\{ 
(\varphi,\vartheta)\in\left[0,2\pi\right]\times\left[0,\pi\right) \left.\right| 
\vartheta(\varphi)\leq \vartheta_{\rm u}(\varphi)) \right\}$.

\begin{figure}[t]
	\centering
		\includegraphics[width=0.45\columnwidth]{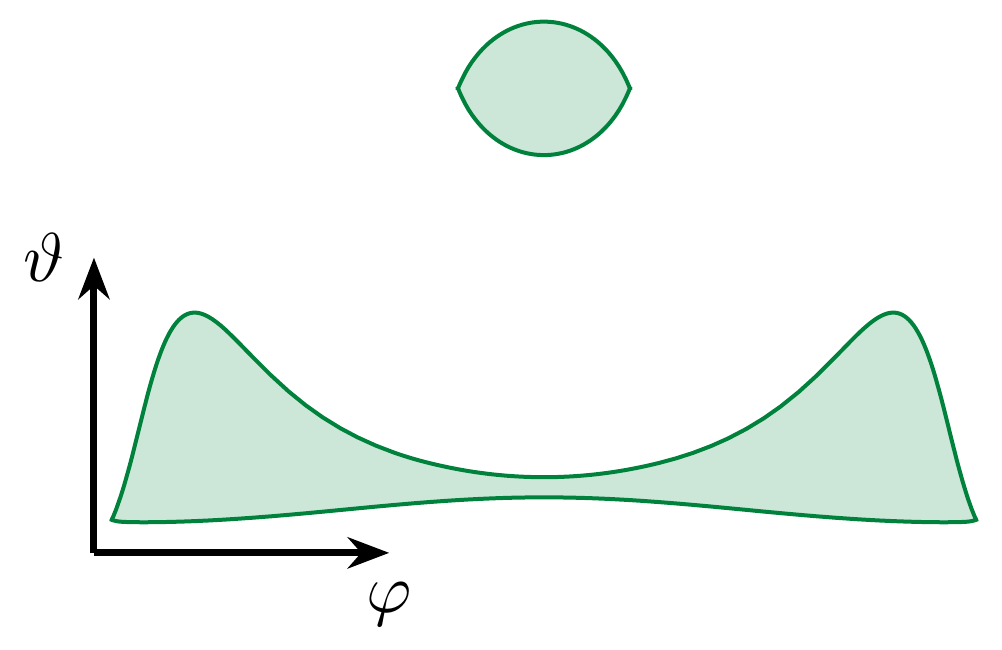}
	\includegraphics[width=0.45\columnwidth]{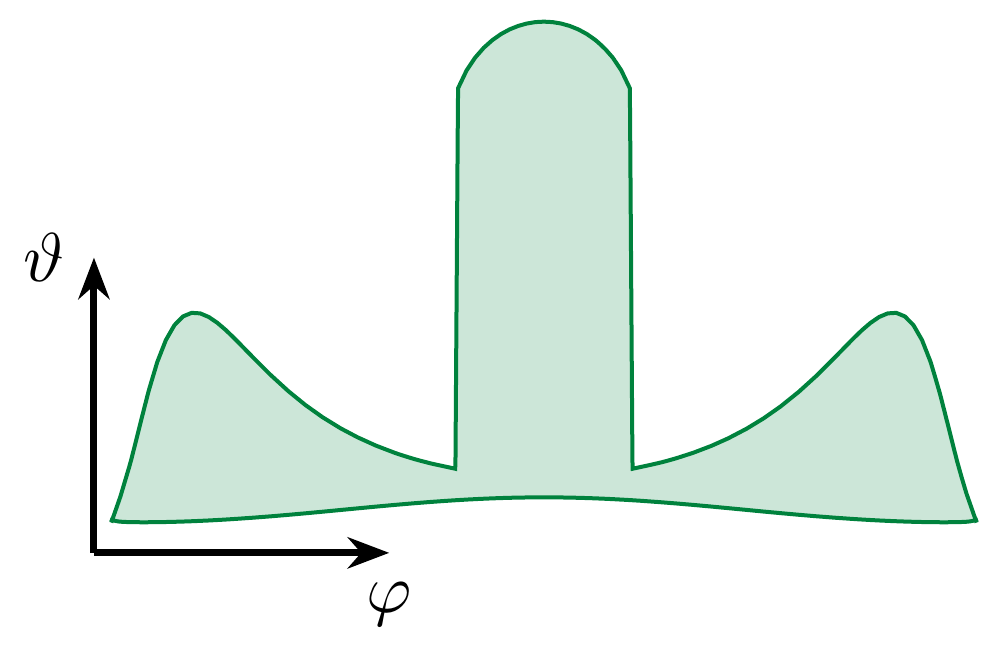} 
	\caption{Potential source of error of the algorithm for determining the 
solid angle region of the dominant signal. The frames show the range where 
$\rho^{\rm Sig}_{\bar{\mu}}(\varphi,\vartheta) = \rho^{\rm 
Bg}_{\bar{\mu}}(\varphi,\vartheta)$. Left is the real requested region shown; on 
the other hand, right is the result according to the used numerical method 
shown. Since the numerical method always assumes exactly one upper and one lower 
bound, the multiple mappings of $\varphi$ to $\vartheta$, as visible left, 
cannot be detected. Such a case was observed at a collision angle 
$\vartheta_{\rm col}=90^{\circ}$ and circular focusing $\mu=5$ of the probe 
beam.} 
	\label{fig:Error_RegionA}
\end{figure}

The iterative algorithm may provide us with an erroneous result for 
$\mathcal{A}_{\rm d,\bar{\mu}}$. 
Figure \ref{fig:Error_RegionA} illustrates a potential source of error: if for 
a fixed value of $\varphi$ more than two values of $\vartheta$ exist, then the 
numerical method presented above fails. 
However, it nevertheless makes sense to adopt this method for the following 
reasons: 
first, this algorithm remains fast and compact in comparison to additionally 
resolving distinguished regions with multiple values of $\vartheta$ for fixed 
values of $\varphi$.
Second -- and most importantly --  we search for regions fulfilling 
$\int_{\mathcal{A}}\!{\rm d}\Omega[\rho^{\rm 
Sig}_{\bar{\mu}}(\varphi,\vartheta)-\rho^{\rm 
Sig}_{\bar{\mu}}(\varphi,\vartheta)]>0$.
As long the total integral over $\mathcal{A}$ is positive, the signal dominates 
the background, regardless of whether this is also the case locally in 
$\mathcal{A}$. A potential failure of the algorithm therefore does not 
lead to wrong estimates for the signal photon number.
Third, we can make sure that region $\mathcal{A}_{\rm d,\bar{\mu}}$ satisfies 
the desired condition by plotting the function 
$\rho^{\rm Sig}_{\bar{\mu}}(\varphi,\vartheta)-\rho^{\rm 
Sig}_{\bar{\mu}}(\varphi,\vartheta)>0$ and compare it with the graph of region 
$\mathcal{A}_{\rm d,\bar{\mu}}$. 

\end{appendix}

\end{document}